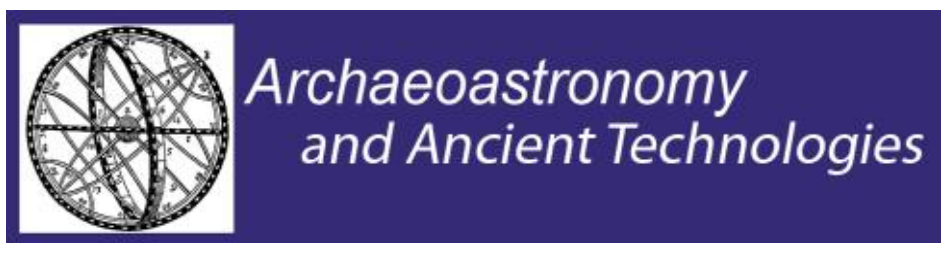

# Arhaeoastronomical Analysis of the Levinsadovka Sacrificial Complex (South Russia)

**Larisa Vodolazhskaya[1,*], Vera Larenok[2]**

[1] Department of Space Physics, Southern Federal University (SFU), str. Zorge, 5, Rostov-on-Don, 344090, Russian Federation: E-mail: larisavodol@yahoo.com

[2] Don Archaeological Society (DAO), str. M. Gor'kogo, 95A, Rostov-on-Don, 344082, Russian Federation; E-mail: dao2@inbox.ru

**Abstract**

The article presents the results of a study of the unique sacrifice of the Levinsadovka sacrificial complex. On it were found large accumulations of animal bones near the buildings. An analogue of this sacrificial complex in the Northern Black Sea region not previously detected. Planigraphy of the sacrificial complex analyzed by arhaeoastronomical methods. The new method of calculating the terrain elevations with the use of topographic maps has been developed within this research to improve the accuracy. The location of bone clusters and the location of single finds of ancient stone tools were compared with the bearings of the rising and setting of the Sun and the Moon in the astronomically significant moments of the year. The orientation to important directions, associated with the Moon, in the organization of the sacrificial complex was identified as a result of this comparison. Detected orientation of the Levinsadovka sacrificial complex allows considering it the Moon sanctuary. Sacrificed fragments of stone tools with smooth surface were located on the Moon directions as did clusters of bones. Using folklore data it is concluded that the stone tools symbolized thunderstones, and were regarded as fragments of the stone Sky or the Moon in the Bronze Age, most likely.

## Introduction

The Northern Black Sea coast is rich in archaeological monuments of different eras. More advanced ancient Mediterranean, Anatolian, Mesopotamian culture influenced on the Northern Black Sea coast through Caucasus and Balkans. For many of them were typical religious constructions oriented on astronomically important directions or at least to cardinal directions. For example, this show dedicated to the study of astronomical regularities of spatial organization: of Egyptian temples and pyramids [1, 2, 3], of building complexes of ancient Egypt [4, 5], of temples of Sicily [6], of Roman temples and cities [7, 8], of ancient Alexandria [9].

In the Northern Black Sea coast arhaeoastronomical studies are still quite small. In Ukraine has been investigated by several Eneolithic mounds with columnar constructions in Lower Dniester, Dnieper and Danube [10], in the Odessa region [11], Srubna culture mounds [12] and the ancient necropolis in the Crimea [13]. In the southern Russia was investigated the Karataevo fortress sanctuary [14, 15]. In all cases were identified astronomical regularities in the investigated archaeological sites. However, a relatively small number of sites, have already been studied, does not allow using astronomical features of spatial organization of religious constructions, as a full historical source. It is therefore important to continue to hold arhaeoastronomical analysis of new archaeological sites of the Northern Black Sea coast.

The Russian-German archaeological expedition discovered in 2009 in the south of Russia, on the coast of Miuss Liman, the unique sacrificial complex of the Bronze Age (Fig. 1, Fig. 2).

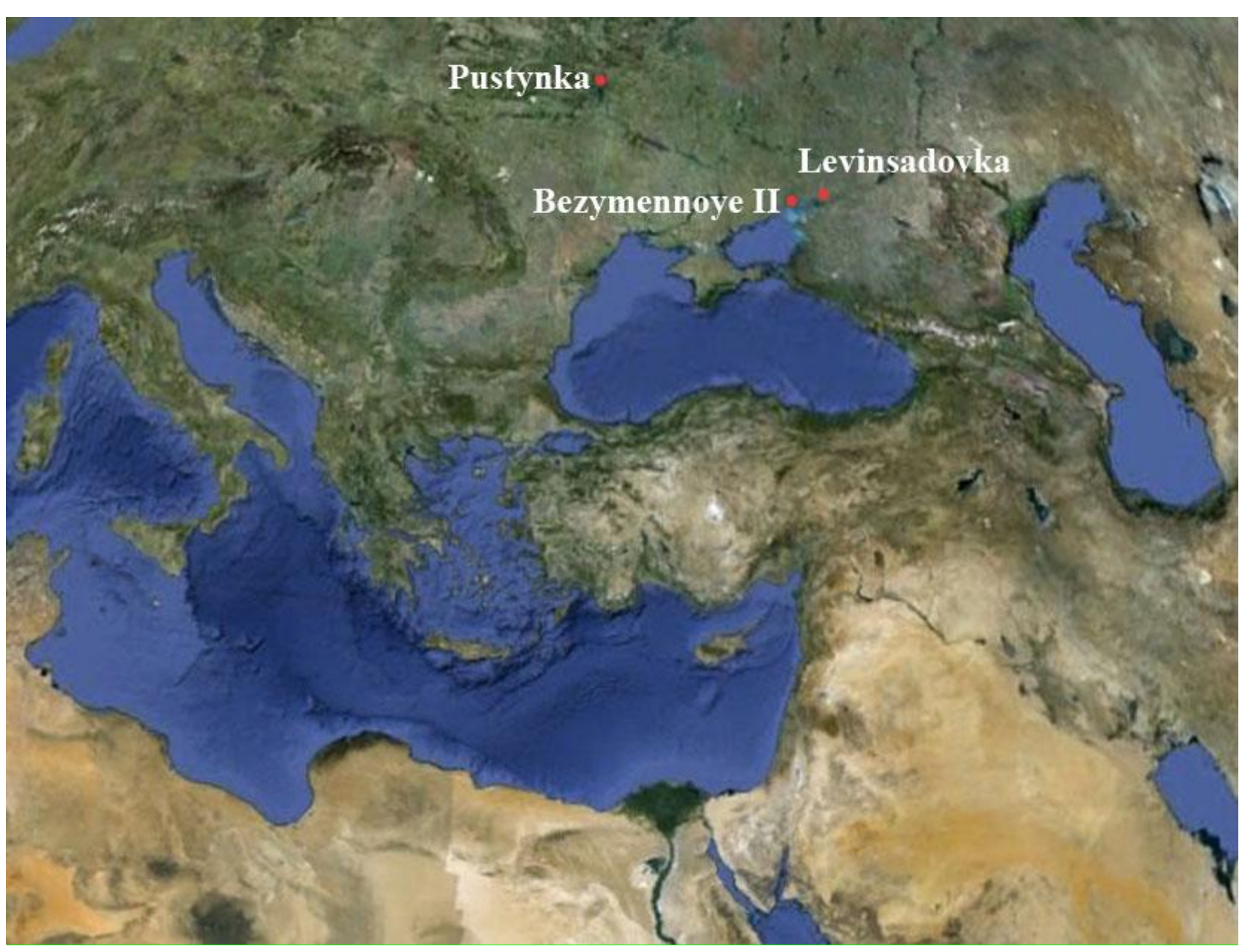

**Figure 1.** The location of archaeological sites

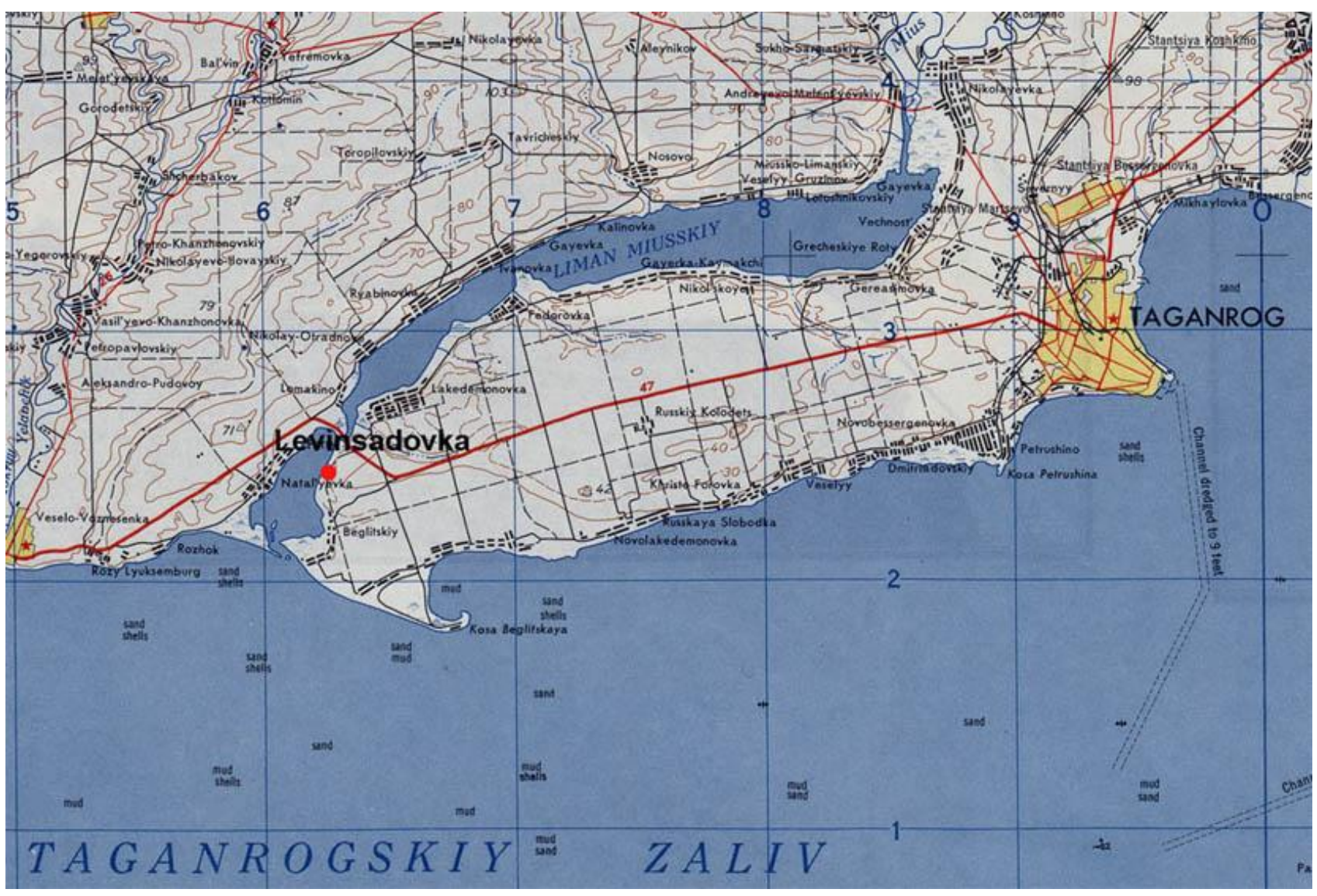

**Figure 2.** Miuss peninsula, Levinsadovka settlement location

We did arhaeoastronomical analysis of this archaeological site, as especially the location of Eurasia ritual structures are often associated with significant astronomical directions, and great sacrifices to be performed during solstices or equinoxes [16].

## Method and results of calculations

We have calculated azimuth of sunrise and sunset at the equinoxes and solstices, as well as high and low azimuth of the Moon for the analysis of astronomical regularities of the Levinsadovka settlement sacrificial complex planigraphy. Azimuth calculation of sunrise and sunset were made by the formulas [17]:

$$\cos A_r = \frac{\sin\delta - \sin\varphi \cdot \sin h}{\cos\varphi \cdot \cos h} \tag{1}$$

$$A_s = 360^0 - A_r \tag{2}$$

where $A_r$ - azimuth of rise, measured from north to east (surveyor), $A_s$ - azimuth of set, $\delta$ - declination, $h$ - altitude, $\varphi$ - latitude. Calculations are made on the upper edge of the disk:

$$h = -R - \rho + p + h_{hor} - h_{cur} \tag{3}$$

where $R$ - 1/2 angular size, $\rho$ - refraction at the horizon, $p$ - horizontal parallax; $h_{hor}$ - angle of terrain elevation on horizon. To account for the curvature of the earth's surface $h_{cur}$ = d × 4.5 × $10^{-6}$, where $d$ - the distance from observer to horizon (horizontal) [18]. At distances up to 15 km us this correction we have not taken into account. For the Sun, the Moon $R$=16$^{/}$, $\rho$=35$^{/}$ [19].

$$p = \arcsin\frac{r}{l} \tag{4}$$

where $r$=6.378 × $10^6$ m - distance from the Earth center to observer on the Earth surface, $l$ - average distance from the Earth center to celestial body center. For the Sun $l$=1.496 × $10^{11}$ m, $p$=8.8$^{//}$. For the Moon $l$=3.844 × $10^8$ m, $p$=57$^{/}$ [20].

During summer solstice the Sun declination equal to angle of ecliptic inclination to celestial equator ε, which is calculated using the formula:

$$\varepsilon = 23.43929111^0 - 46.8150^{//} \cdot T - 0.00059^{//} \cdot T^2 + 0.001813 \cdot T^3 \tag{5}$$

$$T \approx \frac{(y - 2000)}{100} \tag{6}$$

where $T$ - the number of Julian centuries, that separates this age from noon of the 1 of January 2000, $y$ - year of required age. During winter solstice the Sun declination $\delta$ =-ε, and during equinoxes $\delta$ = *0*. The Moon orbit plane is inclined to ecliptic at angle $i$≈5.145$^0$.

The Major Moon declination in summer solstice $\delta=\varepsilon+i$, in winter solstice $\delta=-\varepsilon-i$, in equinox $\delta=i$. The Minor Moon declination in summer solstice $\delta=\varepsilon-i$, in winter solstice $\delta=-\varepsilon+i$, in equinox $\delta=-i$ [21].

Magnetic declination D was calculated using the program Magnetic declination online calculators (MDOC)[3] with accuracy 30$^{/}$[4]. The program calculates magnetic declination using the model of International Geomagnetic Reference Field (IGRF), intended for the empirical representation of Earth magnetic field.

---

[3] http://www.ngdc.noaa.gov/geomag-web/ (accessed on 15.09.2013)

[4] Google Earth and Google Maps (http://www.sollumis.com/) can produce significant distortion of the magnetic declination and show distorted direction of true north (accessed on 15.09.2013)

For geographical coordinates of Levinsadovka settlement $Lat=47^0 14^/$ N и $Long=38^0 55^/$ E for 2009 calculated magnetic declination $D=7^0 04^/$ E. Magnetic anomaly during excavations on Levinsadovka settlement was not detected. Calculated by formula 5, angle of obliquity of the ecliptic to the celestial equator for 1200 BC $\varepsilon=23^0 50^/ 20^{//}$. Results of our calculations of sunrise and sunset azimuths with formula 1 for astronomically significant events are presented in Table 1.

**Table 1.** Azimuths of the Sun rise/set at equinoxes and solstices; $h$ – altitude, when $h_{hor}=0$, $\delta$ - declination, $A$ - azimuth, $h_{hor}$ - angle of terrain elevation, $A_{tot}$ - azimuth subject to terrain elevation

| phenomenon | $h,^0$ | $\delta,^0$ | $A,^0$ | $h_{hor},^0$ | $A_{tot},^0$ |
|---|---|---|---|---|---|
| summer solstice, sunrise | -0.85 | 23.84 | 52.31 | 0.84 | 53.15 |
| equinox, sunrise | -0.85 | 0.00 | 89.08 | 0.84 | 89.92 |
| winter solstice, sunrise | -0.85 | -23.84 | 125.40 | 1.18 | 126.58 |
| summer solstice, sunset | -0.85 | 23.84 | 307.69 | 0.82 | 308.51 |
| equinox, sunset | -0.85 | 0.00 | 270.92 | 0.70 | 271.62 |
| winter solstice, sunset | -0.85 | -23.84 | 234.61 | 0.0 | 234.61 |

Results of calculations of azimuths of the Moon rise and the Moon set by the formula 1 are presented in Table 2.

**Table 2.** Azimuths of the major/minor Moon rise/set; $h$ – altitude, when $h_{hor}=0$, $\delta$ - declination, $A$ - azimuth, $h_{hor}$ - angle of terrain elevation, $A_{tot}$ - azimuth subject to terrain elevation

| *phenomenon* | $h,^0$ | $\delta,^0$ | $A,^0$ | $h_{hor},^0$ | $A_{tot},^0$ |
|---|---|---|---|---|---|
| northern major standstill moonrise | 0.08 | 28.99 | 44.58 | 0.84 | 45.42 |
| southern major standstill moonrise | 0.08 | -28.99 | 135.66 | 1.01 | 136.67 |
| northern minor standstill moonrise | 0.08 | 18.69 | 61.94 | 1.01 | 62.95 |
| southern minor standstill moonrise | 0.08 | -18.69 | 118.26 | 1.18 | 119.44 |
| northern major standstill moonset | 0.08 | 28.99 | 315.42 | 0.56 | 315.98 |
| southern major standstill moonset | 0.08 | -28.00 | 224.34 | 0.0 | 224.34 |
| northern minor standstill moonset | 0.08 | 18.69 | 298.06 | 0.84 | 298.90 |
| southern minor standstill moonset | 0.08 | -18.69 | 241.74 | 0.0 | 241.74 |
| equinox major standstill moonrise | 0.08 | 5.15 | 82.49 | 0.84 | 83.33 |
| equinox minor standstill moonrise | 0.08 | -5.15 | 97.68 | 1.03 | 98.71 |
| equinox major standstill moonset | 0.08 | 5.15 | 277.51 | 0.68 | 278.19 |
| equinox minor standstill moonset | 0.08 | -5.15 | 262.32 | 0.58 | 262.90 |

The terrain is very important for arhaeoastronomical research. We propose to use topographical maps to get information about terrain elevations[5].

Distance to the horizon $d_{hor}$ calculated by using the formula 7 (Fig. 3a), which was successfully used in the approximate form in arhaeoastronomical studies already [22]. Curvature of the Earth is not considered at not very large distances.

[5] direction to the true north at each point on the map corresponds to the line of nearest meridian on the map

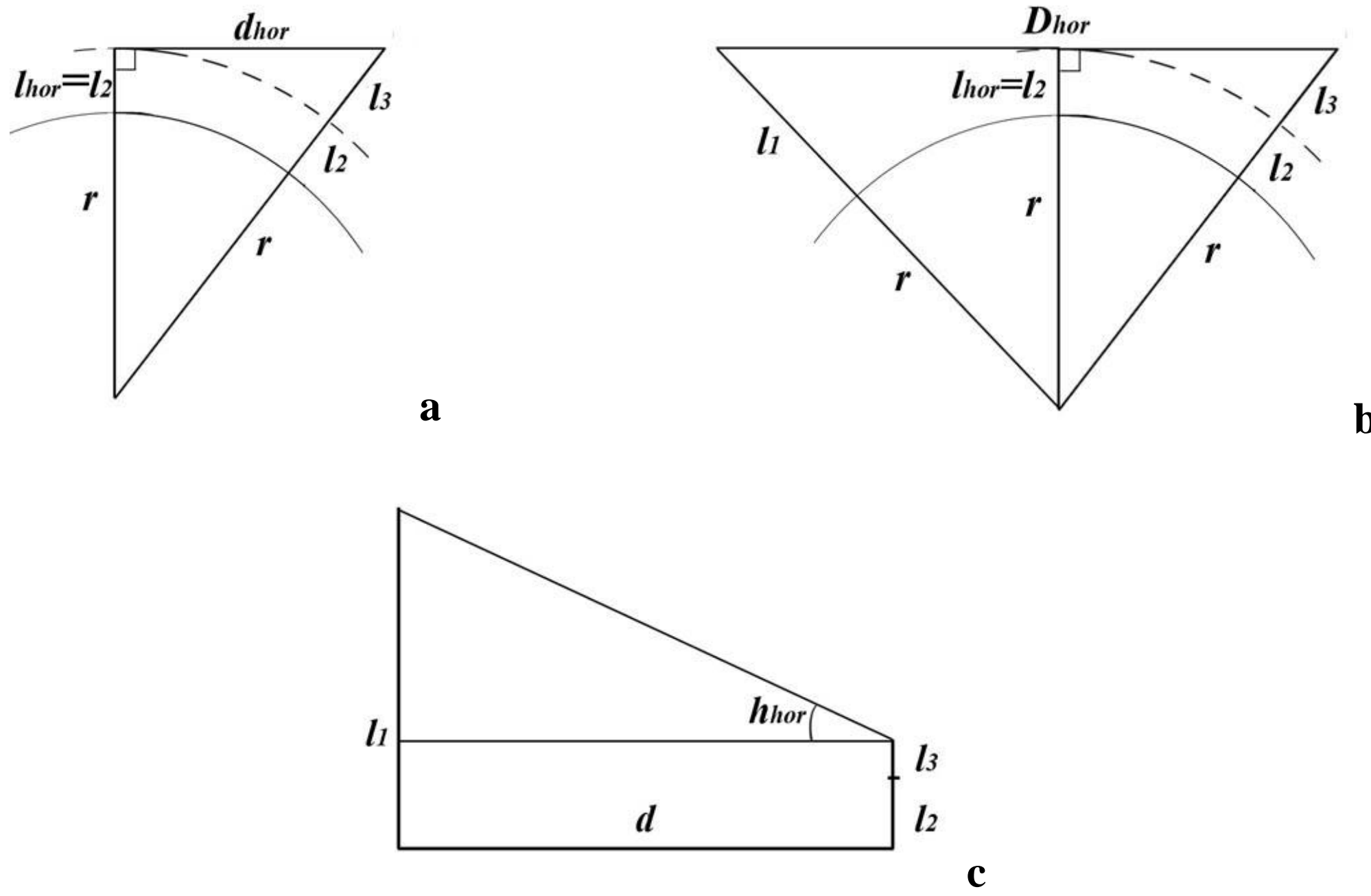

**Figure 3.** Geometrical basis for calculating: **a -** distance to the horizon, **b -** distance of visible above the horizon, **c -** angle of terrain elevation;
$d_{hor}$ - distance to the horizon; $l_{hor}$ - height above sea level on the visible horizon; $l_2$ - height above sea level on archaeological site (view point); $l_3$ - projected growth of the observer; $r$ - distance from Earth center to observer on Earth surface; $D_{hor}$ - distance of visible above the horizon; $l_1$ - height above sea level on the horizontal; $h_{hor}$ - angle of terrain elevation; $d$ - distance from observer to horizontal.

The calculations were performed without refraction, in the approximation that Earth is a sphere[6].

$$d_{hor} = \sqrt{(r + l_2 + l_3)^2 - (r + l_{hor})^2}\ , \quad \text{for } l_{hor} \leq l_2 \tag{7}$$

where $d_{hor}$ - distance to the horizon; $l_{hor}$ - height above sea level on the visible horizon; $l_2$ - height above sea level on archaeological site (view point); $l_3 \approx 1.6$ m - projected growth of the observer; $r \approx 6.378 \times 10^6$ m - distance from Earth center to observer on Earth surface.

Calculated by formula 7 value of distance to the horizon for Azov Sea and Miuss Liman surface, for $l_{hor}$=0 m and $l_2 \approx 4$ m, is $d_{hor} \approx 8452$ m; for flat plain for $l_{hor}=l_2 \approx 4$ m, distance is $d_{hor} \approx 4518$ m.

As altitude increases with distance from the monument, we calculated the distance to terrain elevations visible above the horizon, corresponding to horizontals on topographic map[7] (Fig. 3b).

$$D_{hor} = \sqrt{(r + l_2 + l_3)^2 - (r + l_2)^2} + \sqrt{(r + l_1)^2 - (r + l_2)^2} \tag{8}$$

where $D_{hor}$ - distance of visible above the horizon; $l_1$ - height above sea level on the horizontal "$i$" for $d < d_{hor}$.

Distance of visible above the horizon values for different horizontals, represented on the topographic map in vicinity of Levinsadovka settlement, are presented in Table 3.

[6] contrast the Earth equatorial radius to the polar radius is 0.3%.

[7] For the Levinsadovka used the map «Eastern Europe» 1:250.000, NL 37-2, series N501, U.S. Army Map Service, 1954. http://www.lib.utexas.edu/maps/ams/eastern_europe/ (accessed on 15.09.2013)

If the distance from archaeological site to the horizontal was less than distance to the horizon $d<d_{hor}$, then angle of terrain elevation was calculated for this horizontal by formula 9 (Fig. 3c). Variants of this formula is often used in geodetic leveling [23].

$$h_{hor} = arctg\left(\frac{l_1-(l_2+l_3)}{d}\right), \quad \text{for} \quad d<d_{hor}, \tag{9}$$

where $h_{hor}$ - angle of terrain elevation, $d$ - distance from observer to horizontal, measured on topographic map.

If distance from the observer to the horizontal was $d_{hor}<d<D_{hor}$, then angle of terrain elevation is calculated by the formula 10:

$$h_{hor} = arctg\left(\frac{l_{1i}-l_{1(i-1)}-(l_2+l_3)}{d}\right), \quad \text{for} \quad d_{hor}<d<D_{hor} \tag{10}$$

where $l_{1i}$ - height above sea level on the horizontal with the number "$i$"; $l_{1(i-1)}$ - height above sea level on the horizontal with the number "$i-1$".

The calculations $h_{hor}$ were performed for horizontals in order of increasing height above sea level within distance of visible above the horizon $D_{hor}$ (Tabl. 3, Tabl. 4).

**Table 3.** Distance of visible above the horizon; $l_1$ - height above sea level on the horizontal, $D_{hor}$ - distance of visible above the horizon

| *Levinsadovka* | | *Pustynka* | | *Bezymennoye II* | |
|---|---|---|---|---|---|
| $l_1$, *m* | $D_{hor}$, *m* | $l_1$, *m* | $D_{hor}$, *m* | $l_1$, *m* | $D_{hor}$, *m* |
| 10 | 13266,2 | 110 | 14619,7 | 20 | 15812,0 |
| 20 | 18803,9 | 120 | 19670,7 | 30 | 20490,2 |
| 30 | 22729,1 | 130 | 23416,8 | 40 | 24079,9 |
| 40 | 25947,1 | 140 | 26534,5 | 50 | 27106,2 |
| 50 | 28741,2 | 150 | 29262,4 | 60 | 29772,5 |
| 60 | 31244,8 | 160 | 31718,2 | 70 | 32182,9 |
| 70 | 33533,2 | 180 | 36061,2 | 80 | 34399,6 |
| 80 | 35653,9 | - | - | 90 | 36462,8 |
| 90 | 37639,1 | - | - | 100 | 38400,6 |
| 100 | 39511,8 | - | - | - | - |

**Table 4**. Angle of terrain elevation to astronomically significant directions of Levinsadovka sacrificial complex; $i$ - number of horizontal, $d_i$ - distance from observer to horizontal, measured on a topographic map, $l_{1i}$ - height above sea level on the horizontal, $h_{hor\,i}$ - angle of terrain elevation on the horizontal

| | *phenomenon* | *horizontal* | | | | | | | | |
|---|---|---|---|---|---|---|---|---|---|---|
| | | *i=1* | | | *i=2* | | | *i=3* | | |
| | | $d_i$, *м* | $l_{1i}$, *м* | $h_{hor\,i}$, $^0$ | $d_i$, *м* | $l_{1i}$, *м* | $h_{hor\,i}$, $^0$ | $d_i$, *м* | $l_{1i}$, *м* | $h_{hor\,i}$, $^0$ |
| 1 | summer solstice, sunrise | 300 | 10 | **0.84** | 2000 | 20 | 0.41 | 2900 | 30 | 0.48 |
| 2 | equinox, sunrise | 300 | 10 | **0.84** | 1000 | 20 | 0.83 | 7400 | 30 | 0.19 |
| 3 | winter solstice, sunrise | 200 | 10 | **1.26** | 700 | 20 | 1.18 | - | - | - |
| 4 | summer solstice, sunset | 2600 | 20 | 0.32 | 3500 | 40 | 0.56 | 3800 | 60 | **0.82** |
| 5 | equinox, sunset | 2500 | 20 | 0.33 | 2800 | 40 | **0.70** | 17700 | 60 | 0.14 |
| 6 | winter solstice, sunset | - | - | - | - | - | - | - | - | - |
| 7 | northern major standstill moonrise | 300 | 10 | **0.84** | 3000 | 20 | 0.28 | 17200 | 40 | 0.08 |
| 8 | southern major standstill moonrise | 250 | 10 | **1.01** | - | - | - | - | - | - |
| 9 | northern minor standstill moonrise | 250 | 10 | **1.01** | 2100 | 20 | 0.39 | 3900 | 30 | 0.36 |
| 10 | southern minor standstill moonrise | 250 | 10 | 1.01 | 700 | 20 | **1.18** | - | - | - |
| 11 | northern major standstill moonset | 2700 | 20 | 0.31 | 3500 | 40 | **0.56** | 7300 | 60 | 0.43 |
| 12 | southern major standstill moonset | - | - | - | - | - | - | - | - | - |
| 13 | northern minor standstill moonset | 2600 | 20 | 0.32 | 3300 | 40 | 0.60 | 3700 | 60 | 0.84 |
| 14 | southern minor standstill moonset | - | - | - | - | - | - | - | - | - |
| 15 | equinox major standstill moonrise | 300 | 10 | **0.84** | 1200 | 20 | 0.69 | 6300 | 30 | 0.22 |
| 16 | equinox minor standstill moonrise | 300 | 10 | 0.84 | 800 | 20 | **1.03** | - | - | - |
| 17 | equinox major standstill moonset | 2300 | 20 | 0.36 | 2900 | 40 | **0.68** | - | - | - |
| 18 | equinox minor standstill moonset | 2600 | 20 | 0.32 | 3400 | 40 | **0.58** | - | - | - |

For horizontal, forming the horizon, taken horizontal with maximum angle of elevation. For astronomically significant directions we calculated azimuths $A_{tot}$ by the formula 1 according resulting elevation $h_{hor}$, (Tabl. 1, Tabl. 2) (Fig. 4).

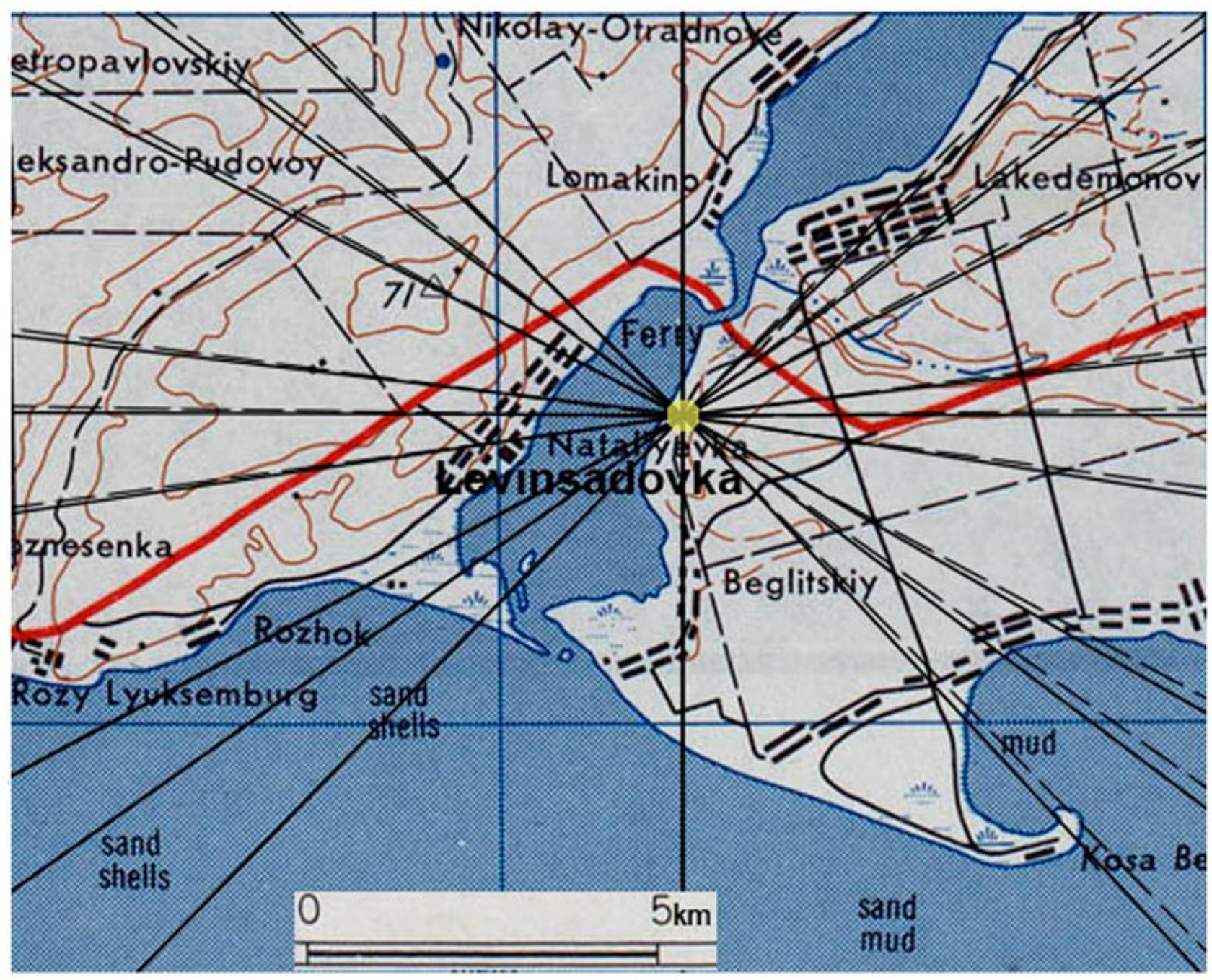

**Figure 4.** The Levinsadovka settlement. Topographic map with applied astronomical directions. Dotted line indicates direction without taking into account of relief elevation

## Object of Study

The archaeological site - Levinsadovka settlement - is located in the western part of the Miuss peninsula on the left bank of the liman in Neklinovsky district of the Rostov region in Russia. Miuss peninsula is a plain, undulating terrain. The settlement is plateau-like area that is elongated in the meridional direction. The settlement is limited by liman to west, from north to the mouth of the beams, from east - the spur of the beam direction, the southern border has no natural boundary (Fig. 5). The Levinsadovka settlement is multi-layer monument founded in the Late Bronze Age.

The Russian-German archaeological expedition led V. A. Larenok and P. A. Larenok (Don Archaeological Society, Russia) and professor Ortwin Dally (German Archaeological Institute, Germany) made archaeological excavation of the Levinsadovka settlement northern section in 2009 [24, 25].

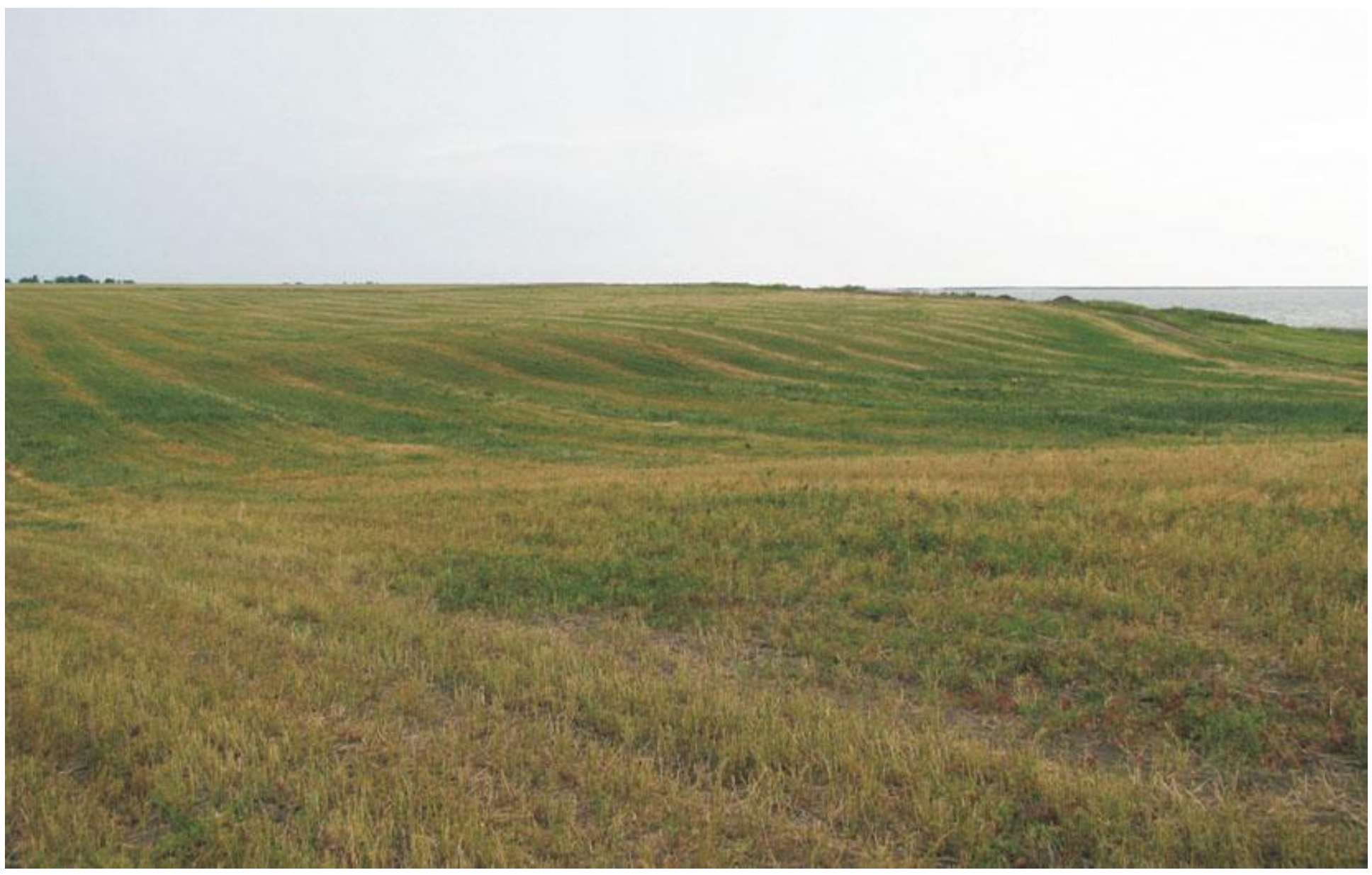

**Figure 5.** The Levinsadovka settlement. View from the North-East [40, Fig. 16]

During the expedition was based two excavation: excavation 1 - at the site of collapse of the indigenous banks of the Miuss liman; excavation 2 - at the northern part of the settlement. In Excavation 2 were found vestiges of two buildings and cultural layers from the Late Bronze Age to Middle Ages. The vestiges of the Late Bronze Age cultural layer are represented by numerous fragments of pottery, flint tools and few stone tools. Particular interests are three fairly large concentrations of animal bones, which were located adjacent to the buildings. Such clusters of animal bones for archaeological sites of the Northern Black Sea coast of the Bronze Age are unique and allow us to interpret the entire complex, as a ritual.

In the cluster 1 demonstrated bone cow (bull home), not less than 3 individuals aged 10-15 years, 4.5 years older and younger than 4.5 years. At the junction underlain bones of spine and ribs, bones of the left front leg cows as part of the carcass and placed on left, along the axis of NE-SW oriented lower back to NE. In addition, there are found isolated bones of sheep, horses, and fish (pike) (Fig. 6).

In the cluster 2 the fossil remains of the bulls home, mainly from 2 - species. The first individual in the age of 1 year and 2-4.5 years older, but here are the bones of other animals of this species in the age of 2.5 years, 2.5 years younger and 3 years older. The second special group by number of fossil remains small livestock (mainly sheep). Quite a large number of sheep bones applies to animals under the age of 1 year. In the same cluster recorded bones of fish - perch, isolated bones of horses and pigs. A few small bone fragments may belong to the human skull (Fig. 7). In the cluster 3 are fragments of bone and whole animal (bull home, sheep, horse), which are dominated by the bones of young bulls. Part of the limb bones of bulls are deposited in the joint (femur and tibia, metapodia). In the southern and eastern part of the cluster marked fragments of skull adult (Fig. 8).

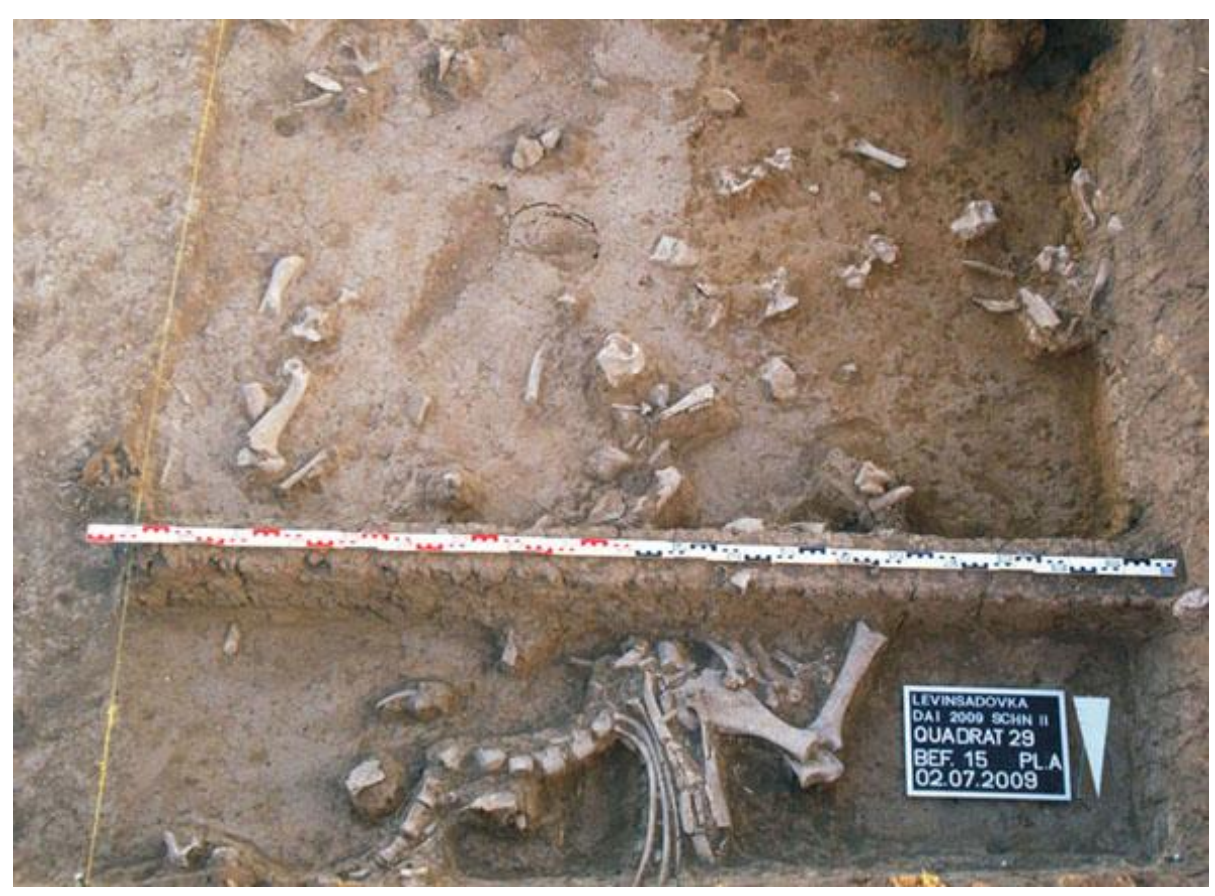

**Figure 6.** The Levinsadovka settlement. Excavation 2. Cluster 1. Plan 7. View from North [40, Fig. 310]

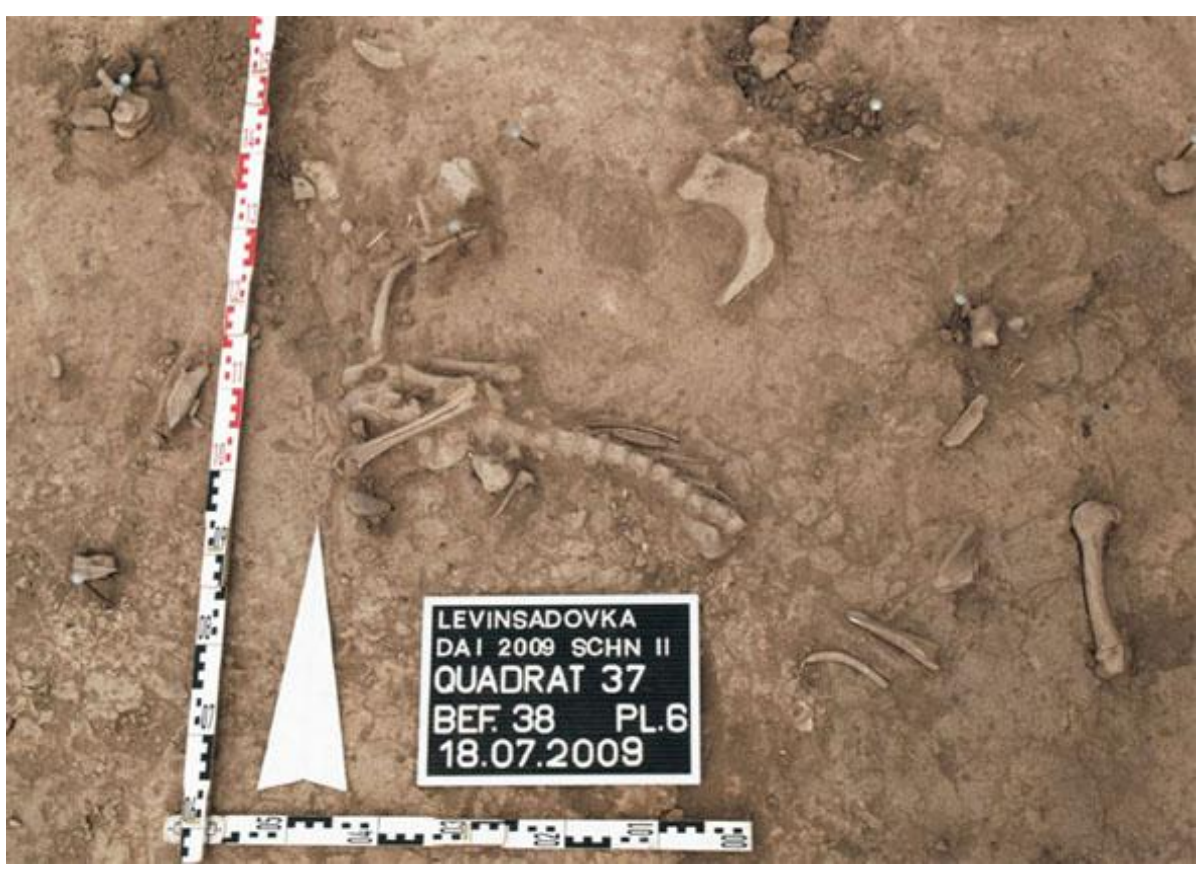

**Figure 7.** The Levinsadovka settlement. Excavation 2. Cluster 2. Plan 6. Fragment. View from South [40, Fig. 315]

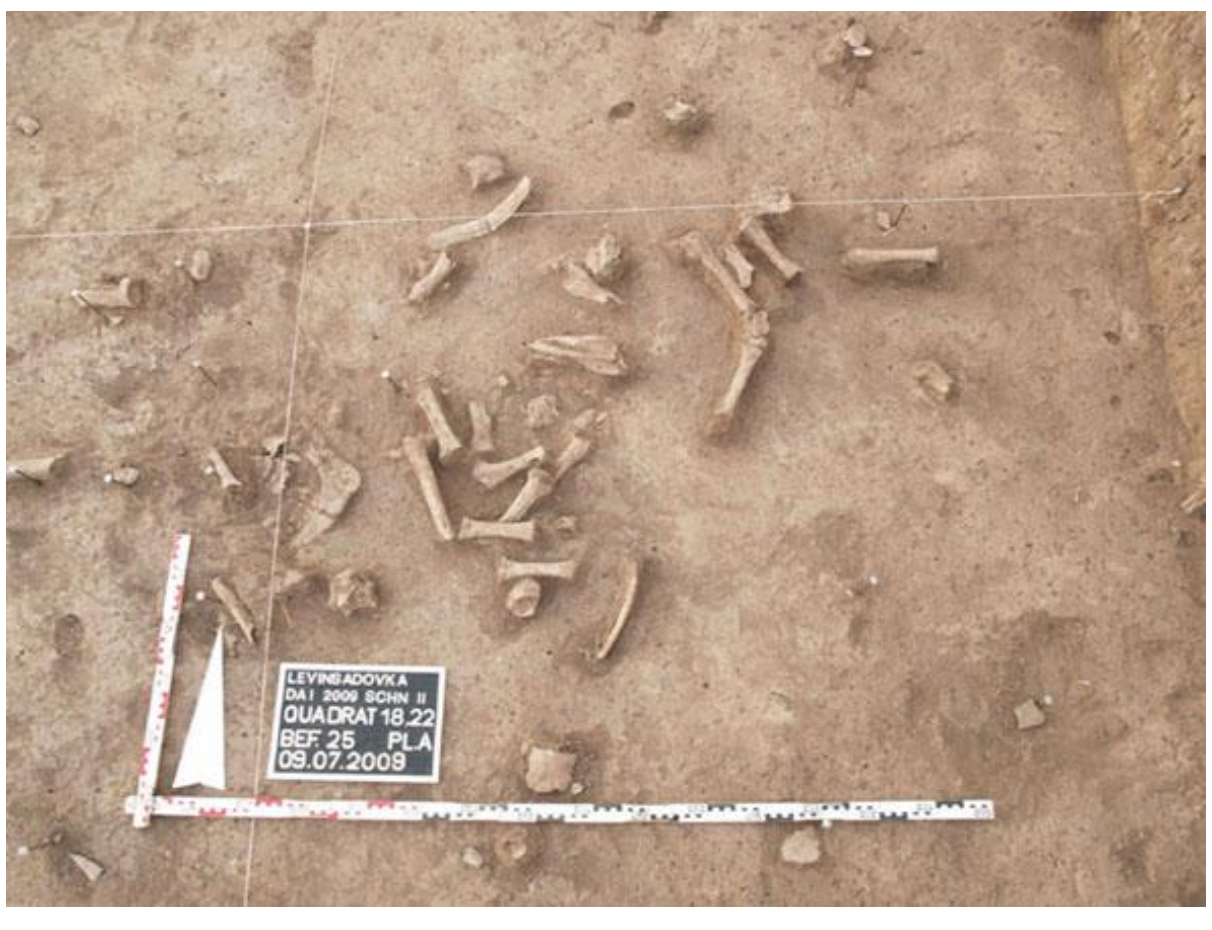

**Figure 8.** The Levinsadovka settlement. Excavation 2. Cluster 3. Plan 10. View from South [40, Fig. 320]

The construction 1, presumably, was the center of the detected sacrificial complex. It was located in the northern part of the Levinsadovka settlement in south-western part of the excavation 2 (Fig.

9). The construction of rectangular construction pit depth in continental clays. The depth of the structure was about 1.60 m.

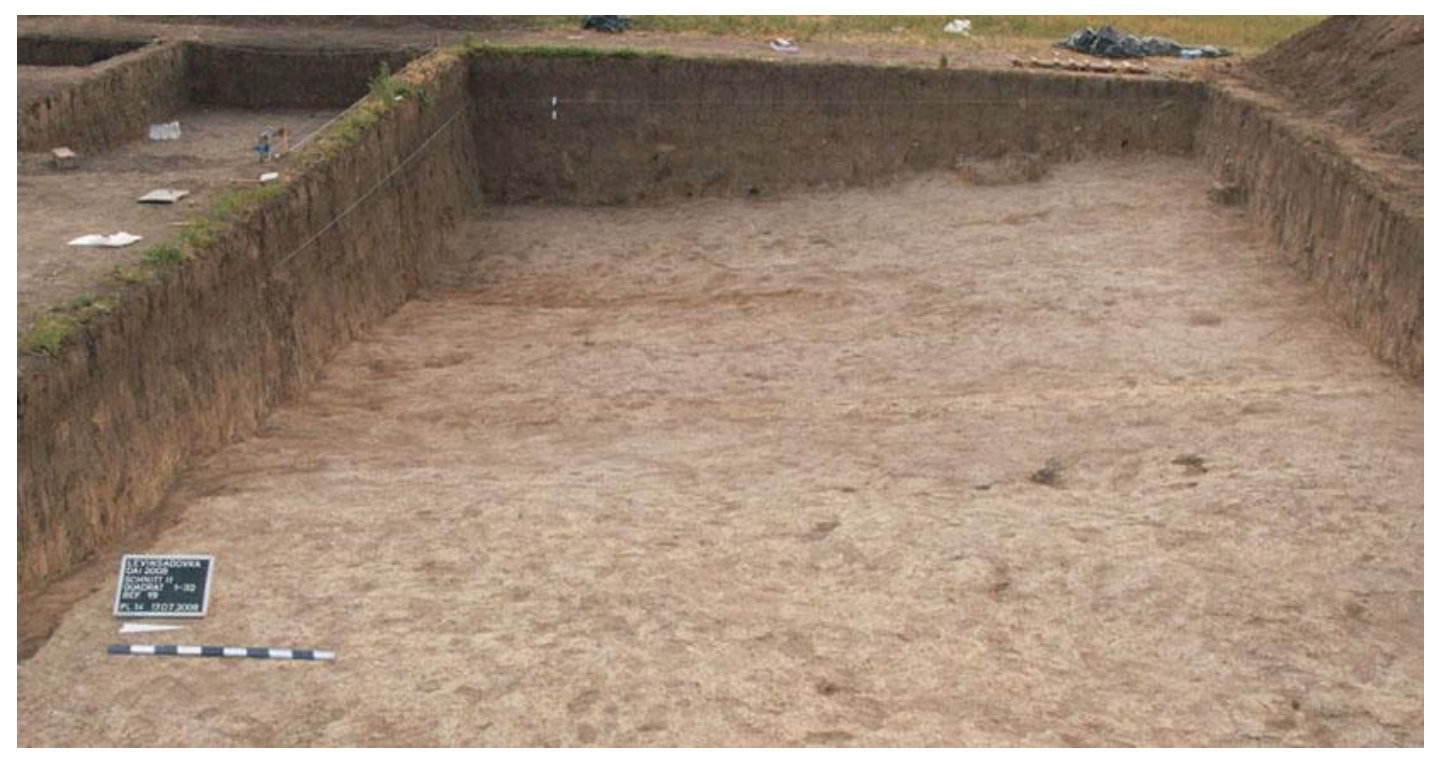

**Figure 9.** The Levinsadovka settlement. Excavation 2. Construction 1. Fragment. The North-Eastern part. View from the east [40, Fig. 293]

Dimensions of the pit to the top of the continental loam is about 7.5 x 5.6 m, the long axis of the pit was located on the north-south line. Western and south-eastern parts of the pit have not been fully traced, as located outside the grid squares of the excavation. In the bottom of the pit were some small lumps of burnt clay coating and small fragments of pottery of the Late Bronze Age.

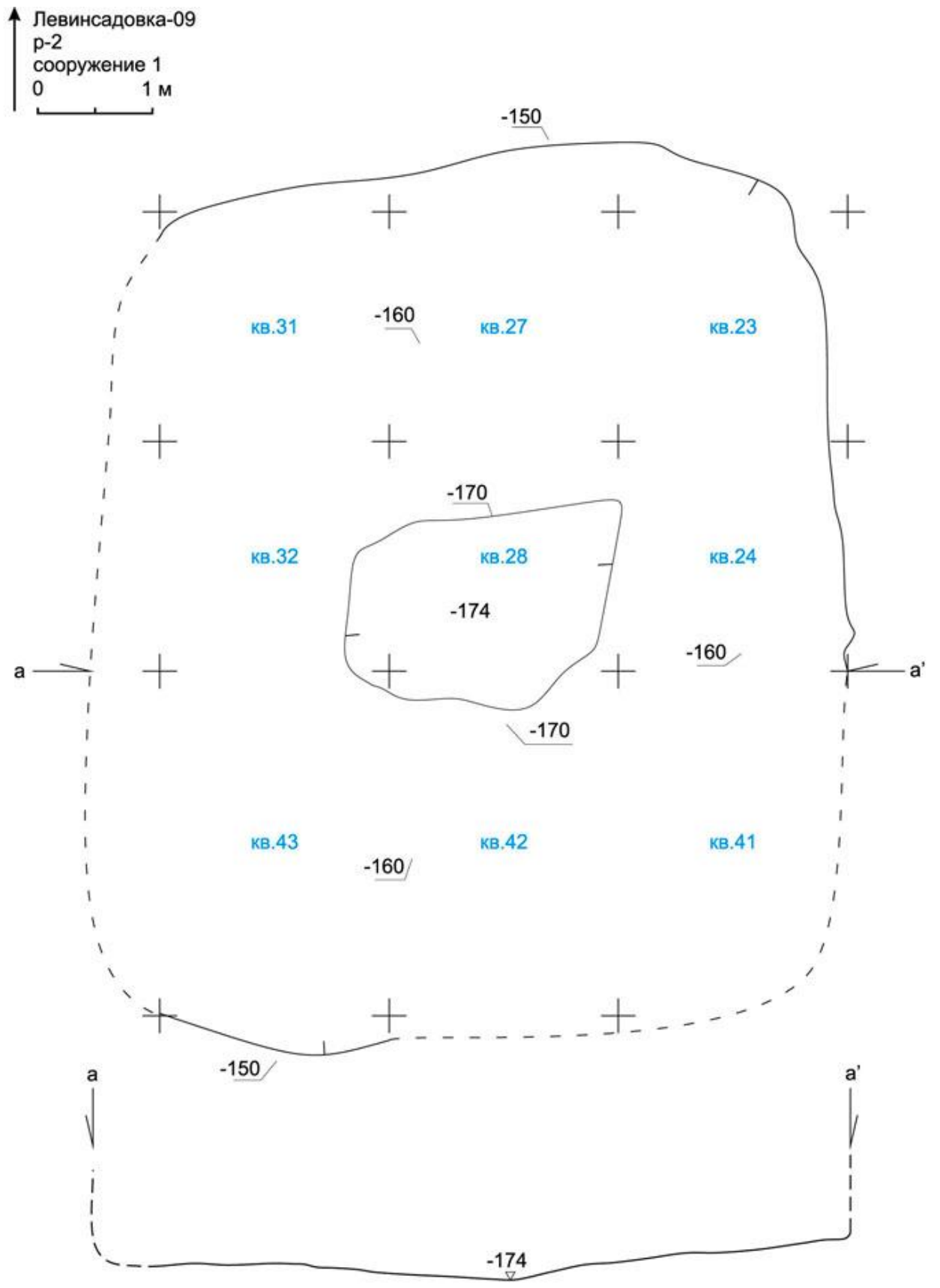

**Figure 10.** The Levinsadovka settlement. Excavation 2. Construction 1. Layout, section. [40, Fig. 299]

In the central part of the pit there were trapezoidal deepening 2.4 x 1.6 m and depth of 0.15 m, long axis oriented along the line W-B (Fig. 10). Most likely, this deepening was location for ritual portable clay brazier. In construction 1 was not detected in either pole-mounted holes or trench from the base of the walls, so it did not have solid walls and roof, and represent only pit dug into ground, possibly with ground part as light canopy.

The construction 2 is located in the north-eastern part of the excavation 2. It was remains of stone structure, possibly, fragment of lower row of dugout pit lining of the Late Bronze Age. Construction dimensions 1.0 x 0.85 m. Rather, the building was dismantled in ancient times.

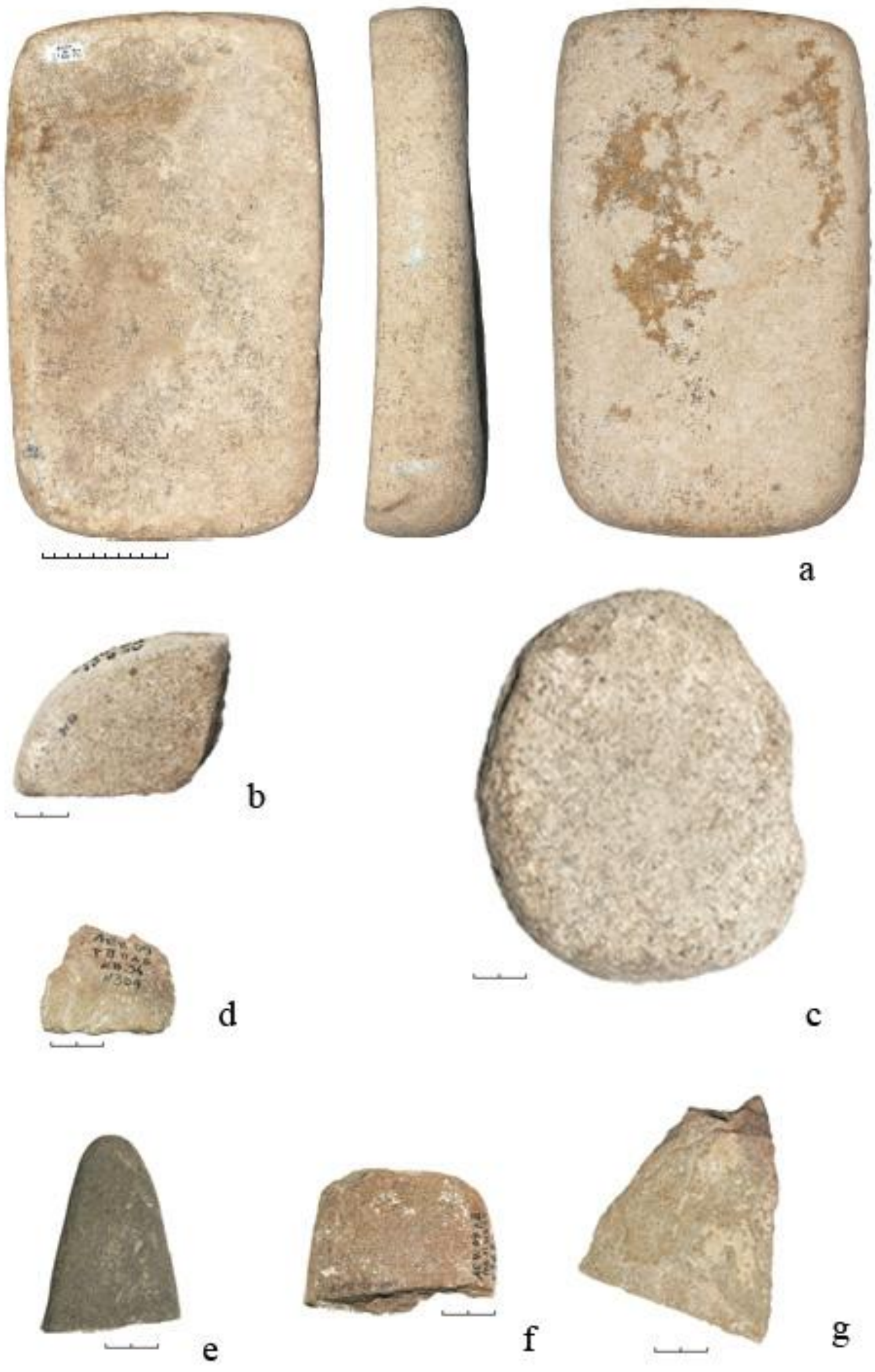

**Figure 11.** The Levinsadovka settlement. Stone tools: **a -** whole millstone (№285) [40, Fig.180-4], **b -** millstone fragment (№471) [40, Fig.281-1], **c -** mortar of gray granite (№571) [40, Fig.180-3], **d -** grinder fragment (№ 304) [40, Fig.180-2], **e -** grinder fragment (№ 350) [40, Fig.226-5], **f -** grinder fragment (№468) [40, Fig.269-14], **g -** grinder fragment (№ 436) [40, Fig.255-3]

Stone tools were found to be unique findings on the Levinsadovka settlement. We believe that they can be used for sacrifice. In excavation 2 was found one whole millstones of gray sandstone tiles with characteristic oval recess in the middle of the size of 37 x 21 x 5.5 cm (Fig. 11a) and a fragment of the same subject (Fig. 11b). For subjects for grinding grains include the small stone mortar of gray granite (Fig. 11c), stones – the grinder of brown sandstone (Fig. 11d), (Fig. 11g) and the stone tool (Fig. 11e). The pest was presented by pebble sandstone fragment with size 7.1 x 5.5 x 2 cm with signs of wear (Fig. 11f).

**Discussion of Results**

The deepening in the construction 1 (the lowest point near the center of the deepening) has been chosen as arhaeoastronomical point of the reference relative to which could perform rituals and sacrifices. The ritual clay brazier probably placed in this deepening. Portable braziers used for the traditional sacrificial ritual – Yajna, which took its origin in the religious practices of the Vedic religion, in the Hinduism so far.

Lines coinciding with most important astronomical orientations were plotted on excavation 2 plan on selected point (Fig. 12). Analysis of elements of the complex showed that cluster 1 is located near true north direction, and cluster 2 and 3 correspond are located on important the Moon directions. Stone tools are in accordance with most important astronomical directions as well. This regularity confirms ritual purpose of complex.

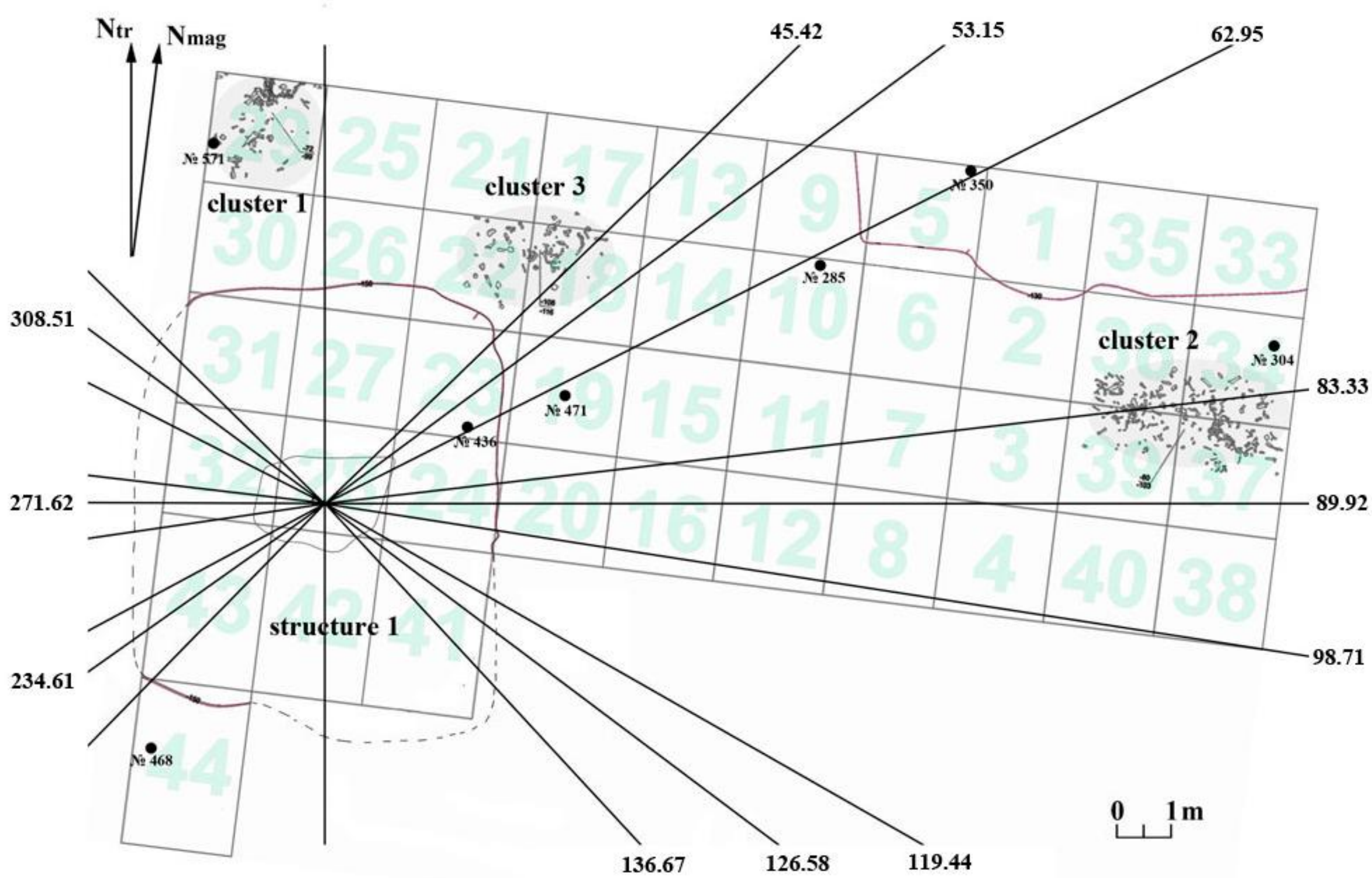

**Figure 12**. The Levinsadovka settlement. Excavation 2 with loop of structure 1 [40, Fig. 282]. On excavation plan: bones clusters, fragments of stone tools, basic astronomical directions with taking into account of relief elevation. Ntr - true North, Nmag - magnetic North.

The Levinsadovka settlements sacrificial complex, including the construction 1 and the sacrificial site around it, was likely part of larger sacred site. About its size is difficult to judge, because archaeological excavations were carried on not very large area. The Levinsadovka settlement population of the Bronze Age belonged to the Srubna culture and to Indo-Iranian ethnic group. It is believed that the ancient Iranians staged sacrificial (altar) areas (no temples) due to nomadic way of life. Strabo reported that the Persians had "pirefies" - large fenced sacred areas with the altar in center [Strabo, XV, 3, 15].

Pirefies are usually located in spectacular high places, on mountains tops, on water body's banks [26]. The Levinsadovka religious complex at is similar in pirefy characteristics and maybe it was pirefy prototype - proto pirefy. The complex is located on a hill - on the plateau-like area on the Miuss liman bank. The fence was around the complex likely. This helped to keep bones clusters of sacrificed animals in compact form.

Herodotus testified that the ancient Persians make sacrifices to gods on high places and withoutdoors [Herodot I, 131]: "Zeus, they usually sacrifice on mountains tops and sky called Zeus. They make sacrifices as Sun, Moon, Fire, Water and Wind. Initially they offered sacrifices only this one deity, and then from the Assyrians and Arabs Persians learned to read Urania". This astral form of ancient Iranians religion, described by Herodotus, it allows connection of spatial orientation of cult activities and sacrifices to apparent motion of celestial bodies.

Correlates bones clusters and astronomical directions relative to deepening the Structure 1 center revealed following pattern. The bones cluster 1 located in immediate vicinity of true North direction, which coincides with projection to earth surface of the world axis - imaginary line, through world center, around which rotation of the celestial sphere. However, Zoroastrians were characterized sacralization South, not North. North for them - evil devas abode. If sacrifices were made to North, it makes them stronger [27]. Only in oldest parts of Avesta, created at the end of II millennium BC [28], direction of North was seen as positive and has been associated with the sacred mountain Hara Berezaiti, around which heavenly bodies. Mountain located as the Indian mythical Meru Mount, in far North, and as Meru is abode of gods. North venerated because of the location of the constellation Ursa Major, this was perceived as the main constellations in the sky [29]. The Soma - Moon is mentioned as a god - the keeper of North side of world in Indian " Manu Laws" [Manu, III, 87].

Thus, the sacrificial animals bones cluster 1 location to North indicates proximity of the Levinsadovka residents cult not to Zoroastrian, and to older - Indo-Iranian system, recognizing sacredness of North with Meru Mount (Hara Berezaiti).

The bones clusters 2 and 3 proved to be connected with the Moon directions. The bones cluster 2 is located in direction of the major standstill moonrise in equinoxes, and the bones cluster 3 - in direction of the northern major standstill moonrise. Location of almost all stone tools correlated with the Moon directions. On the excavation 2 found several stone tools (Fig. 11). The mortar of gray granite (№ 571)[8] was within the bones cluster 1 (Fig.11c), the grinder fragment (№ 304) was found in the immediate vicinity of the bones cluster 2 and it is synchronous in time (Fig. 11d). Another grinder fragment (№ 468) was found in the neighborhood with the construction 1, near south-western corner (Fig. 11f). He was close to the southern major standstill moonset direction. Another grinder fragment (№ 436) was detected directly in construction 1 to the northern minor standstill moonrise direction (Fig.11g). Nearly on line in same direction, were found: the whole millstone (№ 285) (Fig. 11a), millstone fragment (№ 471) (Fig. 11b), and in construction 2 grinder fragment (№ 350) (Fig. 11e). The central deepening in the construction 1 has asymmetry and

[8] numbers of archaeological finds from the field inventory

elongation in the same direction. This separateness of the last Moon direction definitely shows its great sacred significance.

The bones cluster 3 is earliest sacrifice on the Levinsadovka sacrificial complex. There are bones fragments sheep, ox, horse, and even fragments of human skull. The cluster 3 was recorded at 10 and 11 plans. On these plans recorded the grinder fragment (№ 436) in the construction 1, the grinder fragment (№ 468) near the south-western corner of the construction 1 and the whole millstone (№ 285). The grinder fragment (№ 471) was found near the northeast corner of the construction 1 on earlier plane 12. The grinder fragment (№ 350) was recorded at later plan 8. On even more recent plans to 5 - 7 were recorded bones cluster 1, which shows the bones bull, sheep, horses, and fish. The bones cluster 2, representation of bones bull, sheep, horses, pigs, fish and perhaps small pieces of human skull, were recorded on plans 4 - 7. The stoupa (№ 571), and the grinder fragment (№ 304) corresponding to bones clusters 1 and 2, as well as the whole millstone (№ 285) were recorded on plan 6.

Given the distribution of sacrifices on plans, you can select 4 blocks. The first block is associated with plan 12, the second - with plan 10 and 11, the third - with plan 8, and the fourth - with plans 4-7. Each block had sacrifice millstones or grinders along the line coincident with northern minor standstill moonrise direction. Based on this, we can conclude that the said direction was in the sacrificial complex one of the key.

Only one archaeological site of the Bronze Age in Eastern Europe, which also recorded millstones fragments located approximately on the same line, known so far. This is the Pustynka settlement near Pustynki village of the Chernigov region, Ukraine. It dates by the Bronze Age (XIII - XII centuries BC) and is Sosnitski variant of the Eastern-trzciniec culture (XVI - XI centuries BC), related with the Battle Axe culture (Corded Ware) [30, p. 139]. The highest concentration of Sosnitski monuments observed at confluence of Seym River to Desna River, along Desna course, the Left Bank of Polesye, on Upper Dnieper. Southern boundary of the sites is expected to line Kiev - Romny - Sumi [30, p. 137]. To east from Sosnitski monuments settled Abashevo and Pozdnyakovo archaeological cultures, and to south - Bondarihino and Srubna cultures.

The Pustynka settlement was located on the left bank, in the floodplain of the river, on a sandy hill, between the Dnieper and the lake. The settlement cover about $3 \times 10^4$ $m^2$ [30, p. 15]. Researchers were able identify religious buildings in form of ground structure on the settlement. Movable columnar pits from the building preserved. Cult building was set apart from others homes and away from the bank. Movable columnar pits formed a circle with a diameter of about 8 meters. The form of pits and nature to fill them testified that the movable columnar pits. The deepening of 0,20 - 0,25 m trough-shaped, filled with dark ashes sand, was in the center of cult building [30, p. 76].

Rectangular shape ditch is most important part of the structure. The ditch has depth 0.25 - 0.3 m of a trough-shaped deepening, width of about 0.5 m and length of about 3 m. The ditch was densely filled with 118 stones - fragments of millstones and grinders. Their size ranged from 0.03 - 0.04 m to 0.20 - 0.25 m. All stones were burned, some - a lot. In this case, walls, floor and floor around the ditch had no traces of burning [30, p. 79].

Magnetic declination $D$=$4^0 52^/$ E was calculated for geographical coordinates of the Pustynka settlement $Lat$ =$51^0 25^/$ N и $Long$=$30^0 36^/$ E using MDOC for 1965 (year of excavation beginning). Rise / set azimuths of the major / minor Moon for 1200 BC calculated by the formula 1. Calculated azimuth of the northern minor standstill moonrise A=$59.23^0$, the azimuth of the southern minor standstill moonset A=$238.94^0$.

Maximum angle of terrain elevation $h_{hor}$=0.27$^0$ toward azimuth A=59.23$^0$ observed for horizontal 120 m at distance $d$≈8900 m, with average tree height about 20-25 m [9], with $l_1$=145 m. Corrected azimuth $A_{tot}$ =59.62$^0$ calculated by the formula 1 with terrain elevation (Fig. 13)[10].

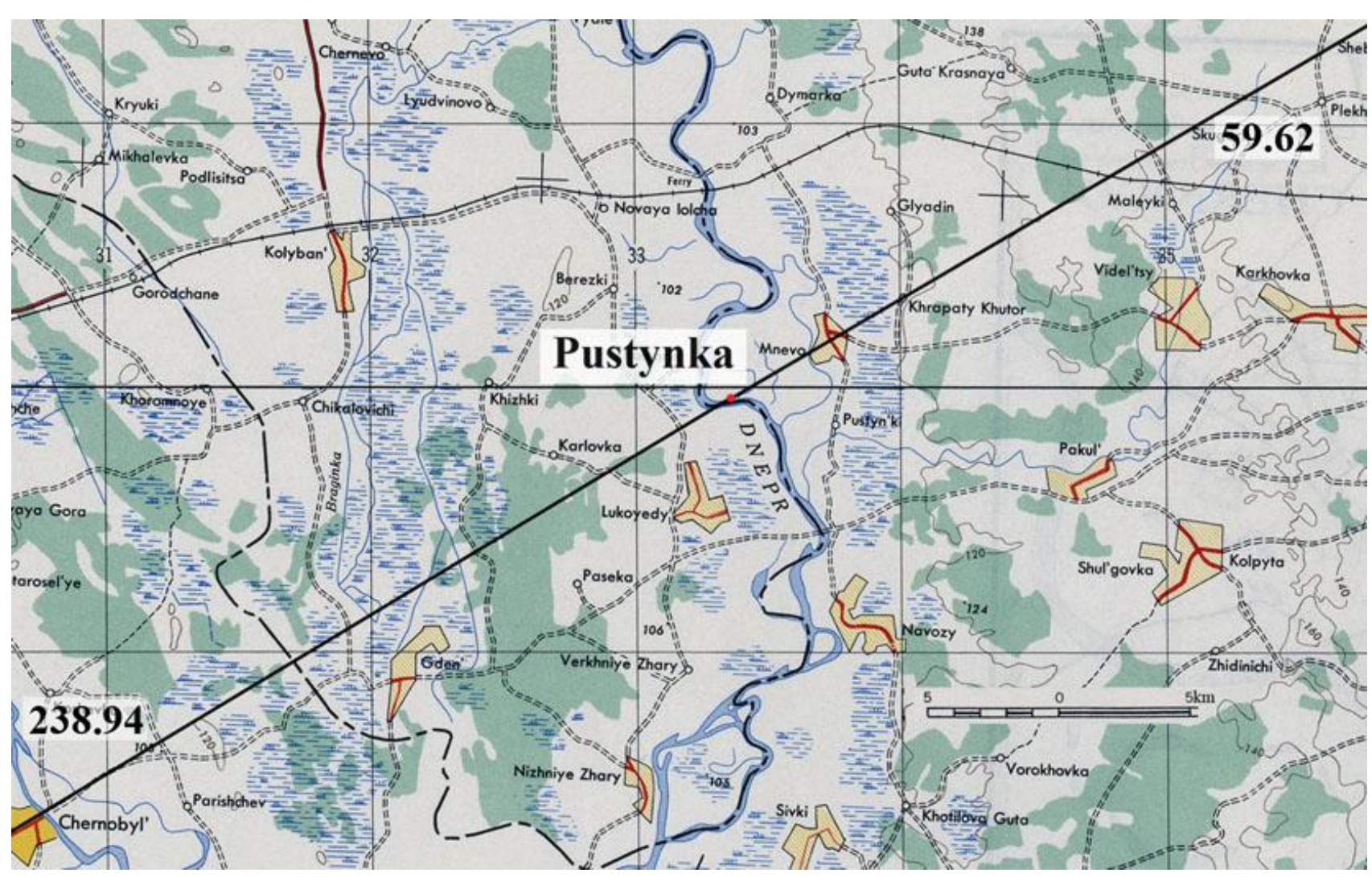

**Figure 13.** The Pustynka settlement. Topographic map with applied astronomical directions: the southern minor standstill moonset direction (238.94$^0$) and the northern minor standstill moonrise direction (59.62$^0$) with taking into account of relief elevation.

Maximum angle of terrain elevation $h_{hor}$=0.1$^0$ toward azimuth A=238.94$^0$ observed for horizontal 120 m at distance $d$≈23700 m, with average tree height about 20-25 m, with $l_1$=145 m. At this distance becomes a significant influence Earth curvature $h_{cur}$=0.1$^0$, so $A_{tot}$ =238.94$^0$.

Approximate center of religious building trough-shaped deepening defined as a point of reference, by analogy with the Levinsadovka construction 1. In this case direction of the ditch will coincide with the southern minor standstill moonset direction (Fig. 14). The southern minor standstill moonset direction and the northern minor standstill moonrise direction are almost identical. This indicates that cult activities in the sanctuary were associated with the minor Moon straight.

The Moon, as the Sun is present in almost all the ancient mythologies. However, neither written nor oral tradition is not brought us legends, would reflect the observations of the major and minor Moon. Reflecting these observations is present in many nations of legends about high and low sky, about waved the sky (with the Moon), and is possible [31]. Many Indo-European nations attended presentation of the stone Sky [32]. Legend exists about how millstone became the Sun, grinder – the Moon and their fragments - stars [33], the tribe Dzhuang in India, which has undergone the Indo-European influence. In the folklore of the modern Indo-European nations such as German, French, there are stories about how the God breaks a hammer the old Moon and makes from their fragments stars [34]. Thus, fragments of stone tools could be perceived in ancient times like fragments of the sky or the Moon.

---

[9] height of trees is taken into account because of the abundance of forest in the area

[10] for the Pustynka used map «Eastern Europe» 1:250.000, NM 36-1, series N501, U.S. Army Map Service, 1947. http://www.lib.utexas.edu/maps/ams/eastern_europe/ (accessed on 15.09.2013)

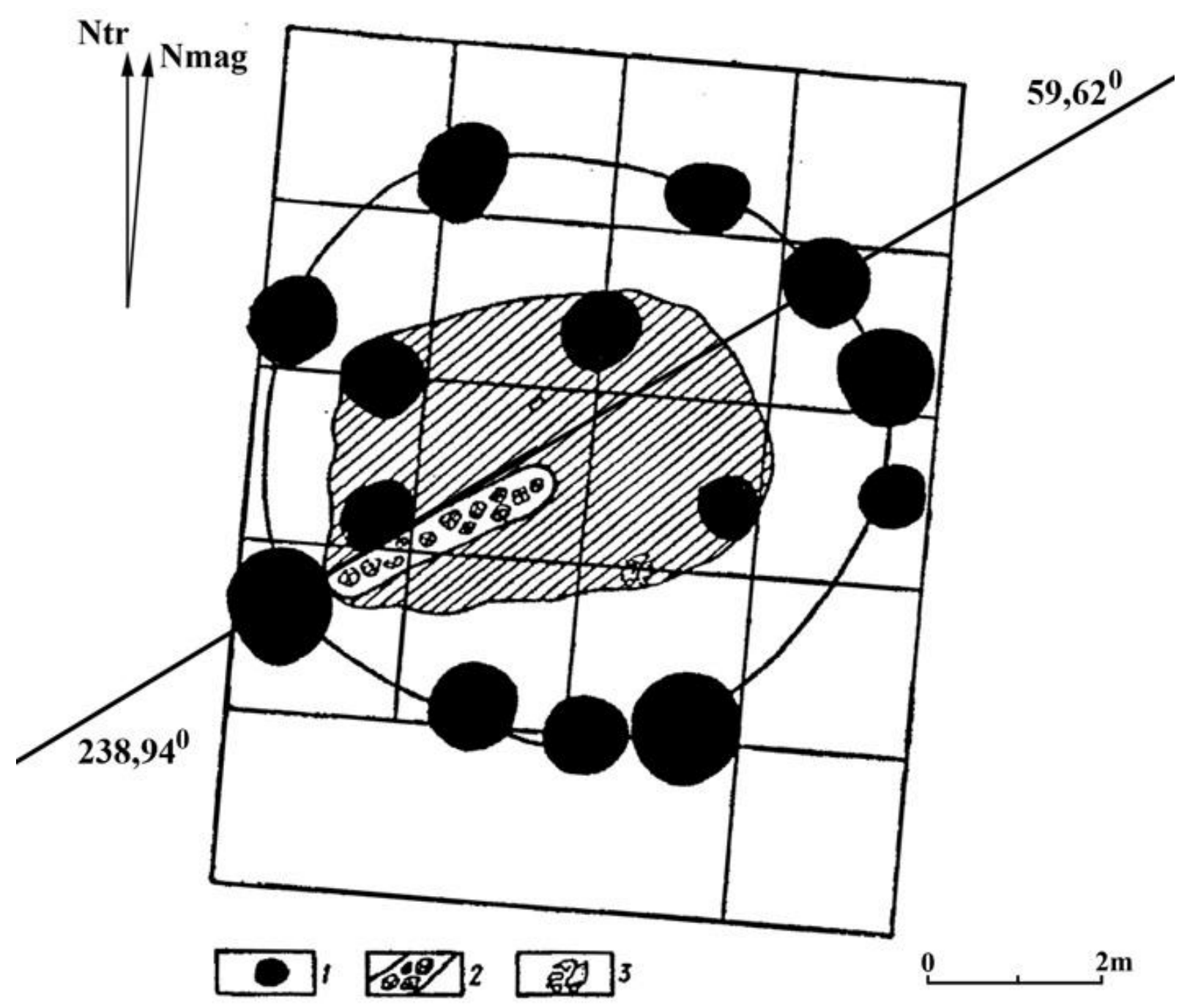

**Figure 14.** The Pustynka settlement. The religious building [30, Fig. 23]. On built plan marked directions to the northern minor standstill moonrise ($59.62^0$) and the southern minor standstill moonset ($238.94^0$) with taking into account of relief elevation. Ntr - true North, Nmag - magnetic North, 1 - holes for posts, 2 – ditch filled by fragments of millstones and grinders, 3 - ceramics

This is confirmed by ancient beliefs about "thunderstones" that allegedly fell from the sky during a thunderstorm and lunar eclipses [35]. It is now established that the "thunderstones" are eneolithic stone axes, flint arrows and various fossils. Many Indo-European nations Thunder God was armed with stone battle ax-hammer initially. Germans have been the Thor's hammer Mjolnir [36], Indo-Aryans had weapons of Indra - the vajra, Lithuanians were sacred the Perkun hammer, and Slavs were the Perun "thunder hammers" [37, p. 252].

In Indo-European languages words are denoting hammer, fracture, ruin and words are denoting flour, mill, millstone go back to the same root "mall". [37, p. 287]. The name of the Thor's hammer - Mjolnir goes back to the same root. In ancient times, originally, same stone was performing functions of millstone, grinder and hammer, maybe. The Bronze Age grinders and millstones nature of similar material and surface treatment with eneolithic stone axes or hammers. Grinders and millstones fragments could replace hammers in some rituals by acting as semiotic signs - models of «thunderstones».

The stone hammer was found in the South sanctuary of Srubna culture on the Bezymennoye II settlement[11]. Hammer was in direction of azimuth of the northern minor standstill moonrise from center of the sanctuary. The South sanctuary was located in the southern outskirts of the settlement, and therefore gets its name [38]. In the central part of sanctuary, at depth of about 1.25 m from current surface and 0.25 - 0.30 m below main surface of the ground in sanctuary center, was fixed large rectangular pit 3.7 x 3.1 x 0.2 m. In the north-western part of which was observed deepening of irregular size of approximately 1.0 x 1.5 m and depth of 0.15 m. In the north-eastern part of

[11] Novoazovsk district of Donetsk region, Ukraine

South sanctuary amongst facing stones, was tethered massive hammer - single stone tool discovered in this sanctuary.

Magnetic declination $D=5^{0}58^{/}$ E was calculated for geographical coordinates of the Bezymennoye II settlement $Lat=47^{0}07^{/}$ N и $Long=37^{0}57^{/}$ E using MDOC for 1996 (year of layout of the South sanctuary excavation)[12]. Rise / set azimuths of the major / minor Moon for 1200 BC calculated by the formula 1. Calculated azimuth of the northern minor standstill moonrise A=62,04$^{0}$, the azimuth of the southern minor standstill moonset A=241,8$^{0}$. Maximum angle of terrain elevation $h_{hor}$=1.2$^{0}$ toward azimuth A=62,04$^{0}$ observed for horizontal 20 m at distance $d\approx400$ m. Corrected azimuth $A_{tot}$ =63.50$^{0}$ calculated by the formula 1 with terrain elevation (Fig. 15)[13].

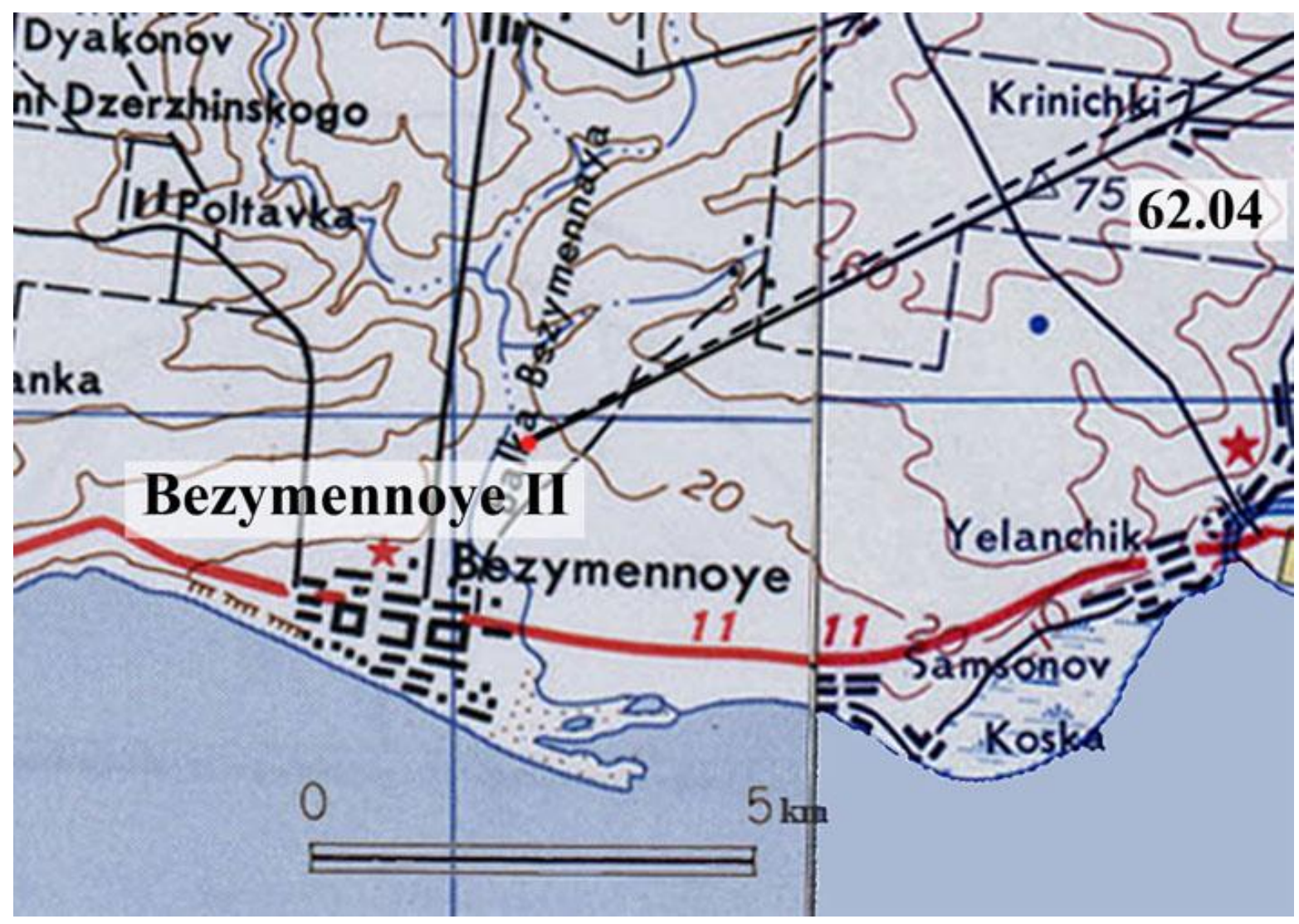

**Figure 15.** The Bezymennoye II settlement. Topographic map with plotted the northern minor standstill moonrise direction (62.04$^{0}$). The dotted line indicates direction without taking into account of relief elevation.

Center of northwest deepening in central square pit was chosen as reference point against which to perform rituals, similar to Levinsadovka. Lunar directions were plotted on excavation plan regarding of the center. The stone hammer was located in the northern minor standstill moonrise direction (Fig. 16).

Location of the South sanctuary stone hammer has same genesis as location of stone tools fragments on Levinsadovka sacrificial complex and in the Pustynka settlement religious building likely. This regularity supports hypothesis that stone tools were playing role of "thunderstones" in rituals of described above sanctuaries.

Fragments of stone tools associated with Moon directions on the Levinsadovka sacrificial complex similar to its smooth surface on the surface fusion of meteorites, so they could also symbolize meteorites. Meteorites associated with the sky in ancient times, like "thunderstones." Ancient iron objects were made of meteoritic iron. Iron is considered a "heavenly" metal in many

---

[12] information of excavation directors V.N. Gorbov and A.N. Usachuk

[13] for Bezymennoye II used map «Eastern Europe» 1:250.000, NL 37-1, NL 37-2, series N501, U.S. Army Map Service, 1954. http://www.lib.utexas.edu/maps/ams/eastern_europe/ (accessed on 15.09.2013)

nations. Ancient Egyptian name of iron "bi-ni-pet" means "heavenly ore" or "heavenly metal." In Ancient Mesopotamia (Ur) iron was called "an-bar" (heavenly metal) [39]. Ancient Greek name of iron "sideros" related ancient Latin word «sidereus», meaning stellar (from «sidus» - star). Fragments of stone tools associated with Moon directions on the Levinsadovka sacrificial complex similar to its smooth surface on the surface fusion of meteorites, so they could also symbolize meteorites.

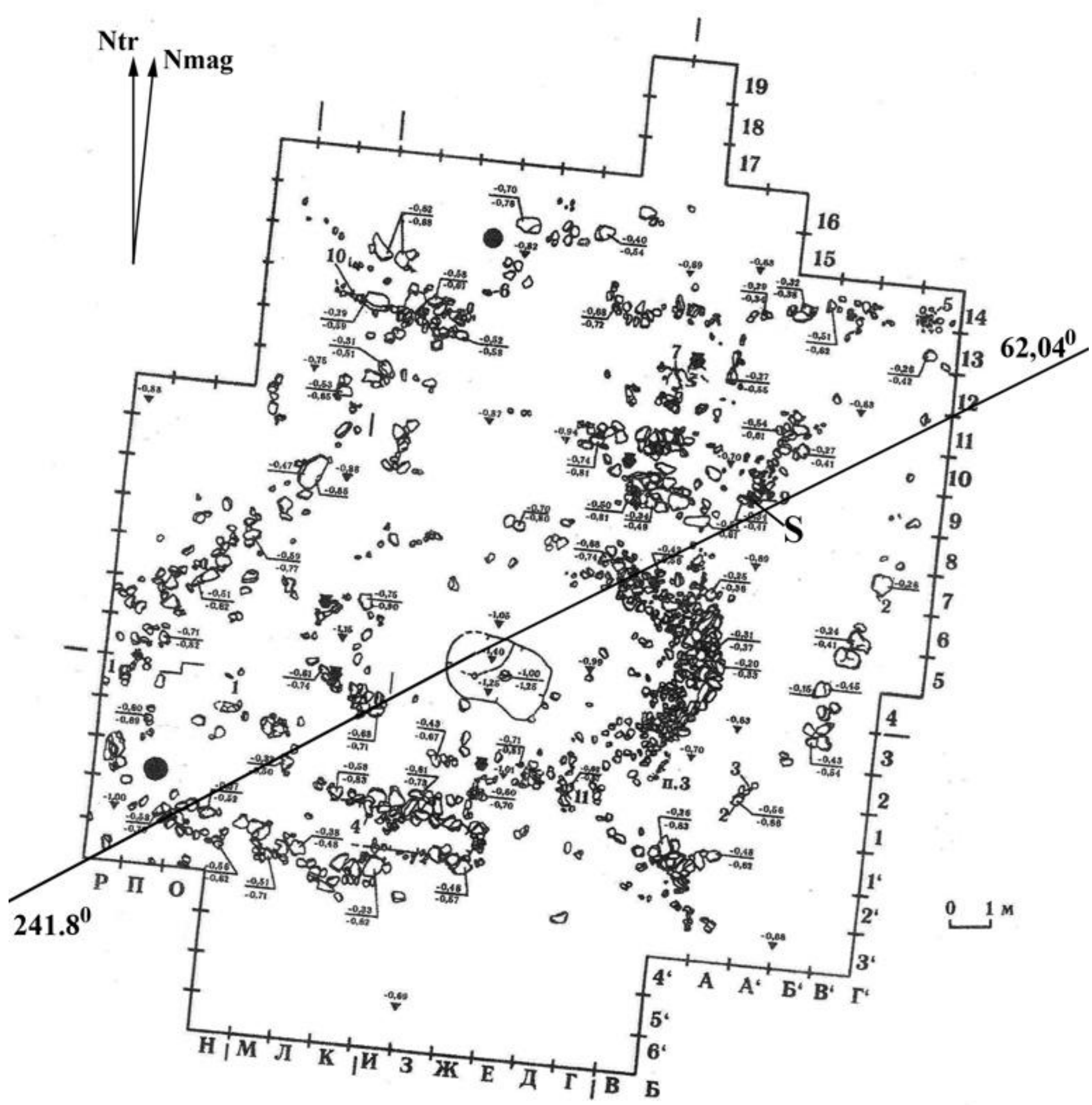

**Figure 16.** The Bezymennoye II settlement. The plan of the South sanctuary [38, Fig. 19]. On the plan marked the northern minor standstill moonrise direction (62.04$^0$) and the southern minor standstill moonset (241.8$^0$) with taking into account of relief elevation. S - location of the stone hammer. Ntr - true North, Nmag - magnetic North

"Thunderstones" in religious activities were semiotic signs - models of "celestial ore" (meteorites) may in turn. The relationship of "thunderstones" with the `northern minor standstill moonrise direction may indicate that point of view on lunar origin of meteorites was common in the Northern Black Sea coast in the Bronze Age.

## Conclusions

Thus, in this study, we have revealed the astronomical regularities in spatial organization of the sacrificial complex. From our point of view, this is evidence of primary astronomical knowledge about the Moon apparent motion of the Srubna population in the Northern Black and of the Eastern-trzciniec population. Arhaeoastronomical methods to reconstruct main principles of spatial organization of the Levinsadovka sacrificial complex and identify it as ancient prototype of pirefy. The special role of north direction will link the of the Levinsadovka sacrificial cult complex with ancient Indo-Iranian tradition. The Levinsadovka sacrifice associated with the worship of deified the Moon, generally. All sacrificial bones clusters and nearly all the fragments of stone tools were located on moon lines relative to center of deepening in cult construction (possible site of a ritual fire). We propose to consider the Levinsadovka sacrificial complex is the lunar sanctuary, as sacrifice of stone tools symbolized "thunderstones" or meteorites, as pieces of the stone Sky or the Moon, from our point of view. We propose the Bezymennoye II settlement South sanctuary and the Pustynka settlement religious building to identify as lunar sanctuaries too. The origin of revealed the Moon cult due to earlier period of the Bronze Age, most likely. The discovery of traces of such cult in early monuments, you'll know region of its origin and ways to further spread.

## References

1. Magli, G. Akhet Khufu: archaeo-astronomical hints at a common project of the two main pyramids of Giza, Egypt, *arXiv*, 2007, http://arxiv.org/abs/0708.3632 (accessed on 15.09.2013).
2. Magli, G. Geometry and perspective in the landscape of the Saqqara pyramids, *arXiv*, 2009, http://arxiv.org/abs/0911.4614 (accessed on 15.09.2013).
3. Sparavigna, A.C. The sunrise amplitude equation applied to an Egyptian temple, *arXiv*, 2012, http://arxiv.org/abs/1207.6942 (accessed on 15.09.2013).
4. Magli, G. A cognitive approach to the topography of the 12th dynasty pyramids, *arXiv*, 2010, http://arxiv.org/abs/1011.2122 (accessed on 15.09.2013).
5. Magli, G. Reconstruction of ancient conceptual landscapes in the Nile Valley, *arXiv*, 2011, http://arxiv.org/abs/1104.1785 (accessed on 15.09.2013).
6. Salt, A. An analysis of astronomical alignments of Greek Sicilian Temples, *arXiv*, 2010, http://arxiv.org/abs/1001.3757 (accessed on 15.09.2013).
7. Magli, G. On the orientation of Roman towns in Italy, *arXiv*, 2007, http://arxiv.org/abs/physics/0703213 (accessed on 15.09.2013).
8. Hannah, R.; Magli, G. The role of the sun in the Pantheon's design and meaning, *arXiv*, 2009, http://arxiv.org/abs/0910.0128 (accessed on 15.09.2013).
9. Ferro, L.; Magli, G. The astronomical orientation of the urban plan of Alexandria, *arXiv*, 2011, http://arxiv.org/abs/1103.0939 (accessed on 15.09.2013).
10. Potemkina, T. M. Osobennosti struktury sakral'nogo prostranstva e'neoliticheskix kurganov so stolbovymi konstrukciyami (po materialam Severnogo Prichernomor'ya) [Features of the structure of sacred space Eneolithic burial mounds with columnar construction (based on the Northern Black Sea)]. *Pamyatniki arxeologii i drevnego iskusstva Evrazii [Monuments of Ancient Art and Archeology of Eurasia]*. M.: Institut arxeologii RAN, 2004, p. 214.
11. Potemkina, T. M. Arxitekturnye osobennosti kurgana Usatovo I-4 [Architectural features mound

Usatove I-4]. *Kratkie soobshheniya Instituta arxeologii. [Brief Communications Institute of Archaeology]. MY. 225,* M., 2011, pp. 206-219.

12. Kirillov, A. Kurgany Doneckoj oblasti kak ob"ekty issledovanij arxeoastronomii [Mounds Donetsk region as research objects archaeoastronomy]. *Arxeologicheskij al'manax [Archaeological almanac].* No. 25. Doneck: «Bytservis», 2011, pp. 180-198.
13. Vinokurov, N.; Macnev, D.; Fesenko, A. Sirius, sozvezdie Bliznecov i syuzhet bozhestvennoj oxoty v Krymskom Priazov'e (na primere nekropolya gorodishha Artezian) [Sirius constellation Gemini and the story of the divine hunting in the Crimean Azov region (for example, the necropolis of the settlement Artesian)]. *Bosporskie issledovaniya. Sbornik nauchnyx trudov. [Bosporus study. Collection of scientific papers.]. MY. XVI,* Simferopol' - Kerch': Krymskoe otdelenie Instituta vostokovedeniya im. A.E. Krymskogo NAN Ukrainy, 2007, pp. 171-190.
14. Vodolazhskaya, L.N.; Vodolazhskij, D.I.; Il'yukov, L.S. Metodika komp'yuternoj fiksacii graficheskogo materiala arxeologicheskix raskopok na primere Karataevskoj kreposti [Fixation technique of computer graphic material of archaeological excavations on the example Karataevskoy fortress]. *Informacionnyj byulleten' Associacii «Istoriya i komp'yuter». [Association Newsletter "History and the computer"].* No 31, M.: Moskovskogo universiteta, 2003, pp. 248-258; http://www.aik-sng.ru/text/bullet/31/248-258.html (accessed on 15.09.2013).
15. Vodolazhskaya, L.N.; Nevskij , M.Yu. Arxeoastronomicheskie issledovaniya svyatilishha Karataevo-Livencovskogo kompleksa [Archaeoastronomical research sanctuary Karataeva Liventsovsky-complex]. *Metodika issledovaniya kul'tovyx kompleksov. [Methods of study of religious complexes].* Barnaul: OOO «Pyat' plyus», 2012. pp. 5-13
16. Potemkina, T. M. Dinamika mirovozzrencheskix tradicij Yuzhnotaezhnogo Tobolo-Irtysh'ya (ot e'neolita do srednevekov'ya). [Dynamics philosophical traditions Tobol and Irtysh yuzhnotaezhnom (from the Chalcolithic to the Middle Ages)]. *Miroponimanie drevnix i tradicionnyx obshhestv Evrazii. [Worldview of the ancient and traditional societies of Eurasia].* M.:Institut arxeologii RAN, 2006. pp. 120-189.
17. Kelley, D.; Milone, E. *Exploring Ancient Skies: An Encyclopedic Survey of Archaeoastronomy.* New York: Springer-NY, 2005, p. 21.
18. Potemkina, T. M.; Yurevich, V. A. *Iz opyta arxeoastronomicheskogo issledovaniya arxeologicheskix pamyatnikov (metodicheskij aspekt). [From the experience of archaeoastronomical studies of archaeological sites (methodological aspect)].* M.: IA RAN, 1998, p. 19.
19. Abalakin, V.K. *Astronomicheskij kalendar'. Postoyannaya chast'.* [*Astronomical calendar. Permanent part*]. M.: Nauka, 1981, p. 44.
20. Montenbruk, O.; Pfleger, T. *Astronomiya na personal'nom komp'yutere. [Astronomy on the personal computer].* SPb.: Piter, 2002, p. 33.
21. Thom, A. *Megalithic lunar observatories*. Oxford University Press, 1971, 127 p.
22. Magli, G. Topography, astronomy and dynastic history in the alignments of the pyramid fields of the Old Kingdom. *Mediterranean Archaeology and Archaeometry*, 2010, Vol. 10, No. 2, pp. 59-74.
23. Red'kov, V.S. *Rukovodstvo po texnicheskomu nivelirovaniyu i vysotnym teodolitnym xodam. [Technical Manual leveling and high-altitude traverse moves].* M.: Nedra, 1974, p. 31.
24. Larenok, V.A.; Larenok, P.A. Raskopki poseleniya Levinsadovka v Neklinovskom rajone Rostovskoj oblasti v 2009 g. [Excavations in settlements Levinsadovka Neklinovsky district of Rostov region in 2009]. *Istoriko-arxeologicheskie issledovaniya v g. Azove i na Nizhnem Donu v 2009 g. [Historical and archaeological research in Azov and the Lower Don in 2009*]. MY. 25, Azov: Izdatel'stvo Azovskogo muzeya-zapovednika, 2011, pp. 78-96.
25. Hoof, L.; Dally, O.; Schlöffel, M. Staying Home or Staying with your Cattle? Different Reactions to

Environmental Changes in the Late Bronze Age of the Lower Don Area (Southern Russia). *eTopoi. Journal for Ancient Studies*. Special Volume 3 (2012), pp. 71–75; http://journal.topoi.org/index.php/etopoi/article/view/121 (accessed on 15.09.2013).

26. Rapen, K. Svyatilishha Srednej Azii v e'poxu e'llinizma. [Sanctuary of Central Asia in the Hellenistic]. *Vestnik Drevnej Istorii. [Herald of Ancient History]*. No. 4, 1994, p. 132.
27. *Zoroastrijskie teksty. Suzhdenie Duxa razuma (Dadestan-i menog-i xrad). Sotvorenie osnovy (Bundaxishn) i drugie teksty.* [*Zoroastrian texts. Judgment Spirit Mind (Dadestan and menog and hrad). Creation of base (Bundahishn) and other texts*]. M.: Vostochnaya literature, 1997, p. 286.
28. Bongard-Levin, G.M.; Grantovskij, E'.A. *Ot Skifii do Indii. Drevni arii: mify i istoriya. [From Scythia to India. The ancient Aryans: Myth and History]*. M.: Nauka, 1983, p. 167.
29. Campbell, L.A. *Mithraic Iconography and Ideology*. Leiden, 1968, p. 80.
30. Berezanskaya, S.S. *Pustynka. Poselenie e'poxi bronzy na Dnepre. [Pustynka. Bronze Age settlement on the Dnieper]*. Kiev, 1974, 176 p.
31. Gura, A.V. Lunnye pyatna: sposoby konstruirovaniya mifologicheskogo teksta [Lunar spots: ways of constructing a mythological text]. *Slavyanskij i balkanskij fol'klor. Semantika i pragmatika teksta. [Slavic and Balkan folklore. Semantics and pragmatics of the text].* M.: Indrik. p. 461.
32. Pisani, V. II paganenismo balto-slavo. *Storia delle religioni*. vol 2. Torino, UTET: 55-100 (orig. ed. 1949), 1965, p. 842.
33. Verrier, E. *Myths of Middle India*. Madras: Oxford University Press, № 12, 1949, p. 114.
34. Krappe, A. H. *La Genese des Mythes*. Paris: Payot, 1938, p. 111.
35. Zanda, B.; Rotaru, M. *Meteorites: Their Impact on Science and History*. Cambridge University Press, 2001, p. 20.
36. Dubov, I. V. O datirovke zheleznyx shejnyx griven s priveskami v vide «molotochkov Tora». [On the dating of iron neck hryvnia with pendants in the form of a "hammer of Thor"]. *Istoricheskie svyazi Skandinavii i Rossii IX-XX vv. [Historical ties Scandinavia and Russia IX-XX centuries].* L.: Nauka, 1970, p. 262.
37. Afanas'ev, A. N. *Poe'ticheskie vozzreniya slavyan na prirodu. Opyt sravnitel'nogo izucheniya slavyanskix predanij i verovanij, v svyazi s mificheskimi skazaniyami drugix rodstvenyx narodov. [Poetic views on the nature of the Slavs. Experience a comparative study Slavic traditions and beliefs in connection with mythical tales of other related peoples*]. T. I, M.: Izd. K. Soldatenkova. 1865. 803 p. (http://books.google.ru/books?id=iakOAAAAQAAJ&printsec=frontcover&source=gbs_atb#v=onepage&q&f=true) (accessed on 15.09.2013)
38. Gorbov, V.N., Mimoxod, R.A., Kul'tovye kompleksy na poseleniyax srubnoj kul'tury Severo-Vostochnogo Priazov'ya. [Religious complexes in the settlements carcass culture of the North-Eastern Azov]. *Drevnosti Severo-Vostochnogo Priazov'ya: Sbornik nauchnyx statej. [Antiquities North-Eastern Azov: Collected articles*]. Doneck: Ukrainskij kul'turologicheskij centr, 1999. S. 35.
39. Lippmann, E.O. *Entstehung und Ausbreitung der Alchemie*. Berlin, 1919, p. 612; Bd. Ill, Weinheim, 1954, pp. 57 - 61.
40. Larenok, V.A. *Raskopki poseleniya Levinsadovka v Neklinovskom rajone Rostovskoj oblasti v 2009 godu. Illyustracii k otchetu. [Excavations in settlements Levinsadovka Neklinovsky district of Rostov region in 2009. Illustrations to the report*]. Rostov-na-Donu: NP Yuzharxeologiya. 2010.